\documentclass[final, 5p, twocolumn, times, num]{elsarticle}

\usepackage{amssymb}
\usepackage{amsmath}
\usepackage{bm}

\biboptions{sort&compress}

\journal{Journal of Magnetic Resonance}

\begin{document}

\begin{frontmatter}



\title{Localization-driven exchange contrast in diffusion exchange spectroscopy} 


\author[1]{Teddy X. Cai} 
\ead{teddy.cai@nih.gov}
\author[1,2,3,4]{Nathan H. Williamson}
\author[1]{Peter J. Basser \corref{cor}}
\ead{basserp@mail.nih.gov}
\cortext[cor]{Corresponding author: Peter J. Basser}

\affiliation[1]{organization={Eunice Kennedy Shriver National Institute of Child Health and Human Development},
            city={Bethesda},
            postcode={20894}, 
            state={MD},
            country={USA}}
\affiliation[2]{organization={Military Traumatic Brain Injury Initiative (MTBI$^2$)},
            city={Bethesda},
            postcode={20814}, 
            state={MD},
            country={USA}}
\affiliation[3]{organization={Uniformed Services University of the Health Sciences (USU)},
            city={Bethesda},
            postcode={20814}, 
            state={MD},
            country={USA}}
\affiliation[4]{organization={The Henry M. Jackson Foundation for the Advancement of Military Medicine, Inc. (HJF)},
            city={Bethesda},
            postcode={20817}, 
            state={MD},
            country={USA}}

\begin{abstract}
\small
Diffusion exchange spectroscopy (DEXSY) is a method to probe exchange between domains of varying confinement. Analyses of DEXSY signals typically assume Gaussian diffusion within distinct compartments and first-order exchange kinetics between them. Other situations can yield DEXSY signal contrast with respect to mixing time, however, leading to potentially erroneous interpretation. Here, we demonstrate that a one-dimensional compartment with reflecting boundaries and without relaxation can by itself produce such contrast in certain experimental regimes. The origin of this contrast is the diffusive mixing of spin isochromats initially near versus far from either boundary, as the former can be relatively coherent in an effect known as edge enhancement or signal localization. We consider DEXSY signals in the case of extended field gradients and identical encodings. Signals were generated via a numerical approach that solves the Bloch-Torrey equation in discrete space and time using matrix operators. We find that in the localization regime, an apparent first-order rate constant of exchange, $k$, can be extracted from DEXSY signals even in this minimal system. The measured $k$ is approximately proportional to $D/L^2$, where $D$ is the diffusivity and $L$ is the domain size. Typically, $k \approx \pi^2 D/L^2$. We attribute this localization-driven exchange to the relaxation of spatial magnetization modes with mixing time, noting that $\pi^2 D/L^2$ is the first non-zero eigenvalue of the Laplacian basis. These results demonstrate that DEXSY and related methods such as filter exchange spectroscopy (FEXSY) may not be specific to genuine barrier permeation.
\end{abstract}



\begin{keyword}
\small
diffusion exchange spectroscopy (DEXSY) \sep 
filter exchange spectroscopy (FEXSY) \sep
localization regime \sep 
matrix formalism



\end{keyword}

\end{frontmatter}

\section{Introduction}

Multidimensional NMR methods are a powerful means to characterize heterogeneity and compartmentalization in porous media \cite{Benjamini2020, Topgaard2017, Henriques2021}. Among them is diffusion exchange spectroscopy (DEXSY) \cite{Callaghan2004, Qiao2005}. DEXSY and related methods \cite{Henriques2021, Aslund2009, Lasic2011, Nilsson2013, Lampinen2016, Benjamini2017, Cai2018, Williamson2020, Cai2022, Cai2024} probe exchange or more generally signal transfer between domains of distinct molecular mobility. These methods have been applied to study neural tissue \cite{Ramadan2009, Nilsson2013, Bai2020, BreenNorris2020, Williamson2020, Williamson2023, Li2025}, with the interpretation that they are measuring, or at least are sensitive to, a transmembrane exchange process. 

However, the interpretation of DEXSY contrast as evidence of transmembrane exchange relies on several assumptions. The canonical framework assumes separated compartments with distinct diffusion coefficients (i.e., compartments exhibiting Gaussian diffusion) that are connected by first-order exchange kinetics (see also the K\"{a}rger model \cite{Karger1969, Karger1985} and its extensions \cite{Jelescu2022, Jensen2023}). While this may be an accurate modeling framework in the specific case of barrier-limited exchange between intracellular compartment(s) and relatively free extracellular space, other situations may also lead to DEXSY contrast. 

Consider that central nervous system tissue, gray matter in particular, is characterized by ramified structures that span a broad continuum of length scales \cite{AirdRossiter2026}. Recent theoretical and numerical studies \cite{Khateri2022, Chakwizira2025, Kiselev2026} suggest that ramification alone, i.e., without membrane permeability, can generate exchange-like signatures in experiments similar to DEXSY. This should not be surprising, as the branches of a cellular process constitute domains of varying mobility along the gradient direction just as much as intra- and extracellular space(s).

These findings cast doubt on the specificity of DEXSY and related methods to transmembrane exchange. They also raise an adjacent question: to what extent can restricted diffusion and signal localization generate apparent exchange contrast? It has long been known that in certain experimental regimes, namely strong gradients, the signal profile arising in a restricted domain exhibits ``edge enhancement,'' \cite{Hyslop1991, Putz1992, Callaghan1993, Stepisnik1999, Ozarslan2008} meaning signal near boundaries is more coherent. The theoretical basis for this effect was studied by several authors \cite{Stoller1991, deSwiet1994, deSwiet1995, Hurlimann1995, Frohlich2006, Grebenkov2018, Moutal2019}, and the regime in which it emerges was termed the ``localization regime'' \cite{Hurlimann1995}.

In this regime, the difference in mobility between molecules near versus far from barriers suffices to produce signal contrast in a diffusion NMR experiment. It follows that DEXSY in this regime may be sensitive to \emph{intra}-compartment signal transfer. In this manuscript, we investigate whether DEXSY can detect apparent exchange in a single one-dimensional compartment. That is, in the absence of any other source of signal contrast: e.g., permeable barriers, geometric heterogeneity, and relaxation processes such as surface relaxation. 
 
We find that, indeed, localization alone can yield DEXSY signal contrast. Further, an apparent first-order rate constant of exchange, $k$, can be extracted. It has a typical value of $k\approx \pi^2 D/L^2$, and more generally $k \sim D/L^2$, where $\sim$ denotes approximate proportionality, $D$ is the diffusivity, and $L$ is the compartment length. To understand the effect, consider that signal localization yields non-uniform magnetization profiles. This sensitizes DEXSY to the relaxation of spatial modes --- i.e., the profile projected onto the Laplacian or diffusion operator eigenbasis. The eigenvalue spectrum then determines $k$, with $k$ being typically consistent with the first non-zero eigenvalue, due to it having the slowest decay. Put another way, we find that DEXSY is conditionally sensitive to the spectrum of the diffusion operator. Therefore, DEXSY contrast does not necessarily imply inter-compartmental exchange or exchange between distinct geometric domains.

Let us be more specific about the physical system and NMR experiment. We consider compartment length $L$, where $x \in [-L/2, L/2]$, with reflecting boundary conditions. In general, a DEXSY experiment has two diffusion encoding blocks with gradients applied in the same direction. These are separated by a longitudinal storage period or mixing time, $t_m$. We consider a realization of DEXSY in which the encoding blocks have equal duration $T$, and the gradient amplitude $g$ is constant within the encoding, i.e., each is a constant gradient spin echo (CGSE) \cite{Carr1954}. In terms of the pulsed gradient spin echo (PGSE) \cite{Stejskal1965} nomenclature, the CGSE corresponds to pulse timings $\delta = \Delta$, meaning no separation between gradient lobes. More formally, we assume the effective gradient waveform:
\begin{equation}\label{eq: dexsy g(t)}
    G(t) = \begin{cases}
        +g, & 0 \leq t < \tfrac{T}{2}\\
        -g, & \tfrac{T}{2} \leq t < T \\
        0, & T \leq t < T+ t_m \\
        +g, & T+t_m \leq t < \tfrac{3T}{2} + t_m\\
        -g, & \tfrac{3T}{2} + t_m \leq t \leq 2T + t_m
    \end{cases}
\end{equation}
Radiofrequency (RF) pulses are assumed to be instantaneous. A fit to the decay of the final echo amplitude at $t = 2T + t_m$ with respect to $t_m$ can yield $k$. Note that this realization of DEXSY formed the basis of several prior experimental and theoretical studies \cite{Cai2018, Williamson2020, Williamson2019, Cai2022, Williamson2023, Cai2024, Williamson2025}, and has been shown to be optimal for sensitivity to first-order, two-site exchange \cite{Cai2024, Cheng2023}. 

To constrain the scope, we will ignore relaxation mechanisms (i.e., spin-spin $R_2$, spin-lattice $R_1$, and surface relaxation), and focus on diffusion. Note that these can be addressed to some degree via normalization, see ref. \cite{Williamson2020}. Throughout, we present DEXSY data generated via a numerical method described in the following section. In the Appendices, we consider other approaches to DEXSY data sampling and analysis: namely, the numerical inverse Laplace transform approach (i.e., how DEXSY was originally conceived \cite{Callaghan2004, Qiao2005}) and filter exchange spectroscopy (FEXSY), \cite{Aslund2009}, sometimes called filter exchange imaging (FEXI) \cite{Lasic2011, Nilsson2013}. Importantly, we show that our results also apply to FEXSY and its estimation of a time-dependent apparent diffusion coefficient (ADC). To our knowledge, this is the first investigation of localization effects in the context of DEXSY or indeed double diffusion encoding in general.

\section{Signal generation and fitting approach}

In this section, we describe our approach to generate DEXSY signals and extract $k$. To generate signals, we expand upon the state transition matrix framework described in refs. \cite{Herberthson2025, Cai2025} by interleaving diffusion and gradient-induced phase evolution steps. Conceptually, our approach is similar to that of Callaghan \cite{Callaghan1997}, in that the signal is approximated as a matrix operator product (see also refs. \cite{Caprihan1996, Barzykin1999} and the multiple correlation function framework of Grebenkov \cite{Grebenkov2007, Grebenkov2008}; cf. ref. \cite{Herberthson2017}). In contrast, however, we do not project the magnetization onto the Laplacian eigenbasis, instead evolving it directly on the spatial grid. Similar approaches date to Zientara and Freed \cite{Zientara1980}, with later developments by Blees \cite{Blees1994} and others \cite{Salikhov1996, Sen1999}. 

Consider the initial CGSE encoding. Let the domain $x \in [-L/2, L/2]$ be discretized into $N_x$ bins of width $\Delta x = L/N_x$. Given an initial vector $\mathbf{m}(0)$, i.e., the magnetization at $t = 0$ in each bin, the diffusion process with reflecting or Neumann boundary conditions over a time step $\Delta t$ can be modeled by $\mathbf{m}(t+\Delta t)= \mathbf{A}\, \mathbf{m}(t)$, where $\mathbf{A}$ is the tri-diagonal matrix
\begin{equation}\label{eq: transition matrix}
    \mathbf{A} = 
    \begin{pmatrix}
        1 - p & p &  & \hdots & 0 \\
        p & 1-2p & p & & &  \\
        \vdots & \ddots & \ddots & \ddots & \vdots \\
         & & p & 1- 2p & p \\
        0 &\hdots &  & p & 1-p
    \end{pmatrix}, \quad p\leq 0.5
\end{equation}
and $p$ is the probability of moving to an adjacent bin. Note that the first and last elements of the main diagonal in $\mathbf{A}$ capture the boundary conditions at the pore walls. To reproduce the Brownian mean-squared-displacement in the bulk (i.e., away from boundaries), one has that $2D\Delta t = 2p(\Delta x)^2$, and thus
\begin{equation}\label{eq: p}
    p = {D\Delta t}/{(\Delta x)^2}.
\end{equation}
Throughout, we will assume a uniform initial distribution such that the entries of $\mathbf{m}(0) = \mathbf{1}$ are identical, reflecting the effect of the initial RF excitation.

For a linear gradient along $x$ with waveform $G(t)$, the phase evolution over the interval $[t, t + \Delta t]$ is roughly captured by the factor $\exp\left(\text{i}q(t) x\right)$, where 
\begin{equation}
q(t) = \gamma G(t) \Delta t
\end{equation}
is the \textit{incremental} phase wavevector, and $\gamma$ is the gyromagnetic ratio. That is, we can approximate the waveform as piece-wise constant over the interval. Let $T$ be an even multiple of $\Delta t$. The magnetization vector at the end of the first CGSE encoding can be approximated as
\begin{equation}\label{eq: CGSE product}
\begin{aligned}
    \mathbf{m}(T) \approx &\, \mathbf{Q}(T-\Delta t)\,\mathbf{A}\, \hdots 
   \mathbf{Q}(\Delta t)\, \mathbf{A}\,\mathbf{Q}(0)\, \mathbf{A}\,\mathbf{m}(0)\\
    =&\prod_{n=0}^{T/\Delta t-1}\left[\mathbf{Q}(t_n)\mathbf{A}\right]\mathbf{m}(0),
\end{aligned}
\end{equation}
where
\begin{equation}
    \mathbf{Q}_{ij}(t_n) = \delta_{ij}\, \exp(\text{i}q(t_n)\bar{x}_{j})
\end{equation}
is an $N_x \times N_x$ diagonal matrix that captures phase evolution, $\delta_{ij}$ is the Kronecker delta, $q(t_n) = q(n\Delta t)$, and $\bar{x}_j$ denotes the midpoint of the $j^{\text{th}}$ bin. 

The approximation is valid when displacement per interval is small such that phase evolution may be applied at (intermittent) static positions, i.e., dephasing and diffusion occur in sequence, similar to the narrow pulse approximation \cite{Callaghan1993}. In practice, this requires $p \ll 0.5$ and/or $\Delta x$ to be small compared to the dephasing length $\ell_g = (D/\gamma g)^{1/3}$ \cite{Hurlimann1995}, or the travel distance to accrue $\pi$ radians of phase. The accuracy of each time step could be improved by combining the steps on the half interval $\Delta t/2$ and replacing $\mathbf{A}$ with a Crank-Nicolson procedure \cite{Crank1947}, as described by Sen \textit{et al.} \cite{Sen1999}, though we will show that we can attain sufficient accuracy as written.  

It follows that the magnetization at the end of the DEXSY experiment (see Eq. \eqref{eq: dexsy g(t)}) is approximately
\begin{equation}
    \mathbf{m}(2T + t_m) \approx \prod_{n=0}^{T/\Delta t-1}\left[\mathbf{Q}(t_n)\mathbf{A}\right]  \mathbf{A}^{t_m/\Delta t}\, \mathbf{m}(T),
\end{equation}
assuming $t_m$ is a multiple of $\Delta t$, and noting there is no gradient during $t_m$ ($q = 0$) and that the second encoding has identical waveform to the first. This is convenient as long mixing times can be calculated by raising a sparse matrix $\mathbf{A}$ to some power. Finally, the ensemble DEXSY signal, which we will express solely as a function of $t_m$, is obtained by averaging over all bins,
\begin{equation}
    S(t_m) \approx \frac{1}{N_x}\mathbf{1}^{\mathrm{T}} |\mathbf{m}(2T+ t_m)|,
\end{equation}
where $\mathrm{T}$ denotes the transpose and $|\cdot|$ the modulus. Note that $S(t_m)$ is generally purely real.

An apparent exchange rate, $k$, can then be extracted from a phenomenological three-parameter fit of $S(t_m)$: 
\begin{equation}\label{eq: 3 param fit}
    S(t_m) = \beta_1 \exp{\left(-kt_m\right)} + \beta_3,
\end{equation}
where $\beta_1$ corresponds to the total signal variation w.r.t. $t_m$, and $\beta_3$ is a signal floor extrapolated to $t_m \rightarrow \infty$. This fit assumes that the variation in $S(t_m)$ is due to a first-order rate process that is able to completely dephase the magnetization that undergoes said process. This would be true with barrier-limited exchange and encodings that have strong enough diffusion weighting to dephase any external magnetization. We will show, however, that this fit can also be applied to the system here, yielding $k$ in the absence of such a process.

To conclude this section, we discuss our approach to signal generation more broadly, addressing its formal interpretation and computational cost. Consider that Eq. \eqref{eq: CGSE product} is in essence a discrete solution of the Bloch-Torrey equation:
\begin{equation}
    \partial_t m = D\partial_x^2 m + \text{i}\gamma G(t) x m.
\end{equation}
Briefly, let $W = D\partial_x^2$ and $\Omega(t) = \text{i}\gamma G(t) x$ denote the diffusion and phase operators in continuous space. If we can take $G(t)$ to be piecewise constant, say on the interval $t = [0, T/2]$, then the evolution is
\begin{equation}
    m(t) = \exp\left([W + \Omega]t\right) m(0).
\end{equation}
The Lie–Trotter product formula \cite{Trotter1959} states that given linear operators, here $W + \Omega$, one has that:
\begin{equation}
    \exp\left([W + \Omega]t\right) = \lim_{n\rightarrow \infty}\left[\exp\left(\frac{Wt}{n}\right)\exp\left(\frac{\Omega t}{n}\right)\right]^n,
\end{equation}
meaning that the operators can be applied sequentially over a sufficiently small time step $\Delta t = t/n$. Our approach corresponds to this splitting, with $\mathbf{A}$ and $\mathbf{Q}(t)$ representing approximations of $\exp(W\Delta t)$ and $\exp(\Omega(t)\Delta t)$ in discrete space, respectively. To be more precise about $\mathbf{A}$, it is from the Taylor expansion
\begin{equation}
    \exp\left(\mathbf{W}_{\text{FD}}\Delta t\right) \approx \mathbf{I} + \Delta t \mathbf{W}_{\text{FD}} + O([\Delta t]^2),
\end{equation}
where $\mathbf{I}$ is the identity matrix, and $\mathbf{W}_{\text{FD}}$ is a discrete space, first-order finite difference approximation of $W$; i.e., $\mathbf{W}_{\text{FD}}$ is tri-diagonal with the standard three-point stencil in the bulk of $D/(\Delta x)^2 (1,\,-2,\,1)$, and with the first and last elements of the main diagonal being $-1$ to satisfy the boundary condition. Thus, $\mathbf{A}$ is understood to be equivalent to $\mathbf{A} = \mathbf{I} + \Delta t \mathbf{W}_{\text{FD}}$ --- see again Eq. \eqref{eq: transition matrix} --- and the update rule
\begin{equation}
    \mathbf{m}(t + \Delta t) = \mathbf{Q}(t)\,\mathbf{A}\, \mathbf{m}(t)
\end{equation}
is now understood to be a first-order (with regards to diffusion), Lie-Trotter splitting of the Bloch–Torrey evolution, expressed in a Markov chain or state transition sense. See refs. \cite{Zientara1980, Blees1994, Sen1999} for earlier implementations of the same general idea. 

An advantage compared to Callaghan's matrix formalism \cite{Callaghan1997} is that one avoids Gibbs phenomena (i.e., ringing) due to mode truncation. Regarding computational cost, we note again that $\mathbf{A}$ is sparse such that the number of operations scales as $O(N_t N_x)$ to a constant factor, where $N_t$ is the number of time intervals. This may be compared to $O(N_t N_{\lambda}^2)$ for the matrix formalism, where $N_{\lambda}$ is the number of modes. This suggests comparable cost when $3N_x \sim N_{\lambda}^2$. It should be noted, though, that $N_t$ may be comparatively small for the matrix formalism, while our approach is constrained by $p\leq 0.5$, or equivalently $\Delta t \leq (\Delta x)^2/2D$. Cost aside, one effectively trades off issues of mode truncation for those of space discretization; we believe the latter are generally easier to assess. 

Lastly, we note that the approach is readily extended to other experiments and systems. Arbitrary gradient waveforms can of course be approximated as piecewise constant. Different or additional boundary conditions and higher dimensions may be incorporated by adjusting the entries of $\mathbf{A}$. See the discussion for further details.

\section{Numerical results}

\subsection{Magnetization profiles}

\begin{figure*}
    \centering
    \includegraphics{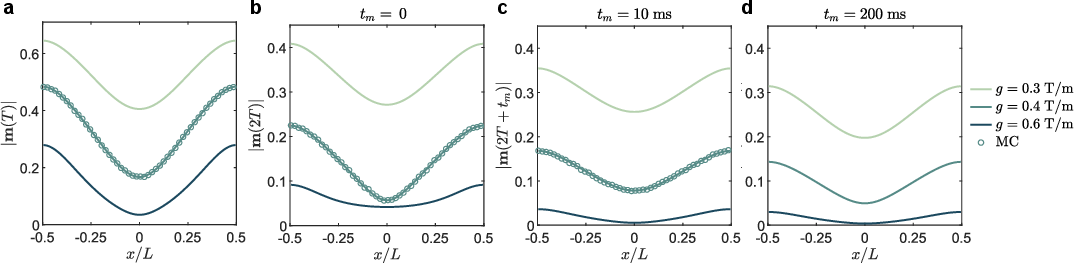}
    \caption{Absolute magnetization vectors $|\mathbf{m}|$ plotted vs. bin midpoints $\bar{\mathbf{x}}$ (solid lines). Parameters were $L = 20\;\mathrm{\mu m}$, $T = 10\;\mathrm{ms}$, $ D = 2\;\mathrm{\mu m^2/ms}$, with varying $g = [0.3,\,0.4,\,0.6] \;\mathrm{T/m}$ (light green to dark blue, respectively), and $\gamma \approx 2.675\times 10^8\;\mathrm{rad/s/T}$. Discretization was $\Delta x = 0.2\;\mathrm{\mu m}$, $\Delta t = 2\;\mathrm{\mu s}$ and $p = 0.1$. Initial condition was $\mathbf{m}(0) = \mathbf{1}$. For $g = 0.4\;\mathrm{T/m}$, MC simulated data is included for comparison, plotted at every other bin (circles). See main text for parameters. (a) Magnetization for the first CGSE encoding block. (b) Magnetization after the second CGSE encoding block, with $t_m = 0$. (c -- d) Keeping the same $y$-axis range as (b), magnetization with $t_m = 10$, $200\;\mathrm{ms}$ between encodings, respectively. }
    \label{fig: profiles vs g}
\end{figure*}

Before looking at more general trends, let us first generate example magnetization profiles to visualize localization and its evolution due to mixing time. In Fig. \ref{fig: profiles vs g}a, the magnitude $|\mathbf{m}(T)|$ generated from Eq. \eqref{eq: CGSE product} is plotted for parameters $L = 20\;\mathrm{\mu m}$, $T = 10\;\mathrm{ms}$, $D = 2\;\mathrm{\mu m^2/ms}$, varying $g = [0.3,\,0.4,\,0.6]\;\mathrm{T/m}$, with $\gamma \approx 2.675\times 10^{8}\;\mathrm{rad/s/T}$ of the proton. To discretize, $\Delta x = 0.2\;\mathrm{\mu m}$ ($N_x = 100$) and $\Delta t = 2\;\mathrm{\mu s}$, yielding $p = 0.1$ from Eq. \eqref{eq: p}. Here it is convenient to introduce the notion of characteristic lengthscales. These are the dephasing length, $\ell_g = (D/\gamma g)^{1/3}$, structural length, $L$, and diffusion length, $\ell_D = \sqrt{DT}$, i.e., the root-mean-square displacement per (effective) gradient application, $T/2$. The localization regime emerges when $\ell_g \ll \ell_D, L$ is the smallest lengthscale. For the parameters above, we have that $\ell_D \approx 4.5\;\mathrm{\mu m}$, and $\ell_g \approx [2.9,\, 2.7,\, 2.3]\;\mathrm{\mu m}$ for the chosen values of $g$. Thus, we expect localization in these profiles, as $\ell_g < \ell_D < L$. Note too that this satisfies $\Delta x \ll \ell_g$, which is important for the validity of Eq. \eqref{eq: CGSE product}, along with $p = 0.1$ being considerably smaller than $0.5$.

For $g = 0.4\;\mathrm{T/m}$, we have included results from Monte Carlo (MC) simulations to validate our approach to signal generation. Simulations were run with the same space discretization, $\Delta t = 10\;\mathrm{\mu s}$, and $3\times 10^5$ walkers. The step size was $\pm\sqrt{2D\Delta t}$. To obtain a profile, a simulation was run for walkers initiated at the center of each bin, $\bar{x}_j$. Each bin was run once. Excellent agreement is observed up to simulation noise, and we proceed with confidence in the accuracy of the signal generation, again given that $p = 0.1$ and $\Delta x \ll \ell_g$. The signal generation was also run with doubled time and space resolution (data not shown) to verify convergence; no meaningful difference was observed.

Note that MC results are plotted versus initial bin position. The generated signal, on the other hand, is in effect plotted vs. final position. Swapping initial and final position is equivalent to a time reversal of the encoding, which in turn is equivalent to reversing the waveform, or $G(t) \rightarrow -G(t)$. As the diffusion operator is real, and the gradient acts via an imaginary phase factor, the two quantities are in fact complex conjugates with identical magnitude that may be compared directly.

In Fig. \ref{fig: profiles vs g}a, one sees that localization is not just a phenomenon associated with strong signal dephasing. Even with relatively weaker gradient amplitudes such as $g = 0.3\;\mathrm{T/m}$, for which $|\mathbf{m}(T)| \gtrsim 0.5$, there is a profile with less dephasing near the boundaries (recall that we initialized with $\mathbf{m}(0) = \mathbf{1}$). In Fig. \ref{fig: profiles vs g}b, the second CGSE encoding block is included, and $|\mathbf{m}(2T)|$ is plotted for the same parameters, with $t_m = 0$. This of course further dephases the magnetization. In addition, it homogenizes the profiles due to the added time and thereby diffusive mixing. This can be seen by comparing the depth of the profiles (from edge to center) --- in Fig. \ref{fig: profiles vs g}a, the depth is $\approx 0.2$, but is closer to $\approx 0.1$ in Fig. \ref{fig: profiles vs g}b.

Following this reasoning, increasing $t_m$ should homogenize the profile further and decrease the overall magnetization as a result. In Fig. \ref{fig: profiles vs g}c, $|\mathbf{m}(2T + t_m)|$ is shown for $t_m = 10\;\mathrm{ms}$, other parameters kept the same. In Fig \ref{fig: profiles vs g}d, that is increased to $t_m = 200\;\mathrm{ms}$. Note that the $y$-axis is kept the same for Figs. \ref{fig: profiles vs g}b -- d to aid comparison. One sees that the profiles are decreased and somewhat more homogeneous, but not uniform. This is because the second encoding can reintroduce signal localization even if $t_m$ is large enough to induce complete mixing --- meaning that at $t = T + t_m$, the magnetization is equilibrated, or nearly uniform. We revisit this notion of equilibration time later, when discussing $k$. Equilibration followed by renewed localization is why Fig. \ref{fig: profiles vs g}d appears to have similar or even slightly less homogeneous profiles than Fig. \ref{fig: profiles vs g}c.      

\subsection{DEXSY signal fits}

\begin{figure}
    \centering
    \includegraphics{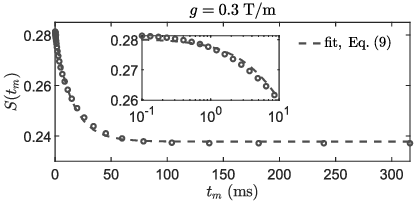}
    \caption{DEXSY signal $S(t_m)$ for the same parameters as in Fig. \ref{fig: profiles vs g}, but just $g = 0.3\;\mathrm{T/m}$. Signals (circles) were generated for 30 values of $t_m$ log-linearly spaced from $10^{-1}$ -- $10^{2.5}\;\mathrm{ms}$, rounded to the nearest multiple of $\Delta t$. A fit of Eq. \eqref{eq: 3 param fit} is shown (dashed line), yielding $k \approx 64\;\mathrm{s^{-1}}$, $\beta_1 \approx 4.3\times10^{-2}$, and $\beta_3 \approx 0.24$, with root-mean-square error $\approx 1.0\times10^{-6}$. The inset shows early decay behavior for $t_m$ up to $10\;\mathrm{ms}$ with log scaling on the $x$-axis. }
    \label{fig: dexsy fit g = 0.3}
\end{figure}

From the trend in Figs. \ref{fig: profiles vs g}b -- d, it is clear that the average magnetization $S(t_m)$ for these parameters will decay such that $k$ can be measured. In Fig. \ref{fig: dexsy fit g = 0.3}, $S(t_m)$ are plotted for $g = 0.3\;\mathrm{T/m}$ and 30 values of $t_m$ log-linearly spaced from $10^{-1}$ -- $10^{2.5}\;\mathrm{ms}$, rounded to the nearest multiple of $\Delta t$. Other parameters were kept the same as Fig. \ref{fig: profiles vs g}. Eq. \eqref{eq: 3 param fit} was then fit to the data, yielding $k \approx 64 \;\mathrm{s^{-1}}$ (this can also be thought of as an apparent exchange time, $1/k \approx 16\;\mathrm{ms}$), $\beta_1 \approx 4.3 \times10^{-2}$, and $\beta_3 \approx 0.24$. While the total signal variation $\beta_1$ is relatively modest, it remains clear that the interaction between localization and mixing time can produce DEXSY signal contrast that appears to be consistent with first-order exchange kinetics. In other words, the decay is roughly monoexponential. The fit is good with a root-mean-square error of $\approx 1.0 \times 10^{-6}$, though we note some systematic bias remains (see inset of Fig. \ref{fig: dexsy fit g = 0.3}), indicating that this is not exactly a first-order process.

We next seek the behavior in various regimes. Specifically, in terms of the characteristic length scales $\ell_D$, $\ell_g$, and $L$, in what regimes does the DEXSY signal from this system exhibit such contrast with $t_m$? Further, what is the degree of contrast (i.e., $\beta_1$), and what, if any dependencies does $k$ exhibit? 

\begin{figure*}
    \centering
    \includegraphics{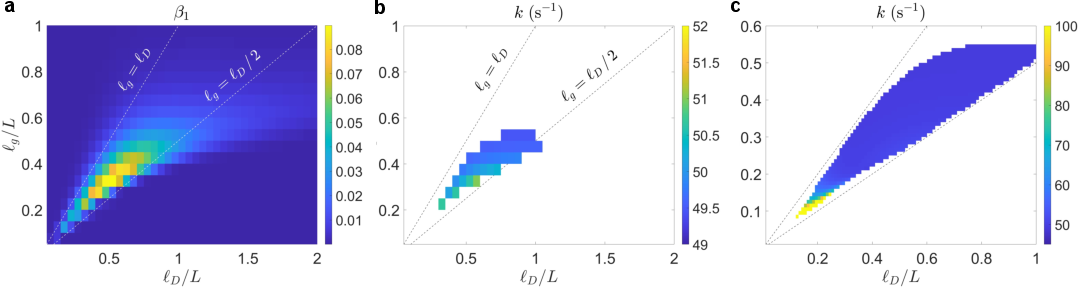}
    \caption{Fits of Eq. \eqref{eq: 3 param fit} for signals generated at $\ell_D$, $\ell_g$ linearly spaced by $0.05L = 1\;\mathrm{\mu m}$, fixing $L = 20\;\mathrm{\mu m}$, $D = 2\;\mathrm{\mu m^2/ms}$, $\Delta x = 0.2\;\mathrm{\mu m}$, and $\Delta t=2\;\mathrm{\mu s}$. The $t_m$ values were the same as in Fig. \ref{fig: dexsy fit g = 0.3}. (a) Total signal variation, $\beta_1$. Detectable signal variation, or $\beta_1 \gtrsim 0.02$, appears in a band lying between the lines $\ell_g = \ell_D$ and $\ell_g = \ell_D/2$ (dashed white lines), up to about $\ell_D \lesssim L$. (b) Corresponding values of $k$ with the same axes as part (a), filtered for $\beta_1 \ge 0.02$. Values are highly uniform $\approx 50 \;\mathrm{s^{-1}}$, see color bar. (c) Values of $k$, focusing only on the region of interest $\ell_D/2 < \ell_g <\ell_D <L$ with finer resolution. The lengthscales $\ell_D$, $\ell_g$ were spaced by $0.01L = 0.2\;\mathrm{\mu m}$. Again, $k$ values shown are filtered by $\beta_1 \ge 0.02$. Note the change in axes range and color scale. }
    \label{fig: param map}
\end{figure*}

To address the first question, signals were simulated for $\ell_D$ and $\ell_g$ linearly spaced by $0.05 L = 1\;\mathrm{\mu m }$, from $0.05L$ up to $2L$ and $L$, respectively. Parameters $L = 20\;\mathrm{\mu m}$, $D = 2\;\mathrm{\mu m^2/ms}$, $\Delta x =0.2\;\mathrm{\mu m}$, and $\Delta t = 2\;\mathrm{\mu s}$ were kept the same as before. That is, we set $g = D/(\gamma \ell_g^3)$ and $T = \ell_D^2/D$. Values of $t_m$ were also kept the same as Fig. \ref{fig: dexsy fit g = 0.3}. The resulting $\beta_1$ values from fits of Eq. \eqref{eq: 3 param fit} are shown in Fig. \ref{fig: param map}a. It can be seen that detectable values of $\beta_1$ --- which we consider somewhat arbitrarily to be $\beta_1 \ge 0.02$, i.e., requiring a signal-to-noise ratio $\gtrsim 50$ --- appear in a narrow, banded region corresponding to $\ell_D/2 \lesssim \ell_g \lesssim \ell_D$ and also $\ell_D \lesssim L$.   

The region can be framed in terms of the characteristic regimes described by H\"{u}rlimann \textit{et al} \cite{Hurlimann1995}. When $\ell_D < \ell_g, L$ is the smallest lengthscale, corresponding to the upper-left of Fig. \ref{fig: param map}a, diffusion is nearly free and there is little localization (i.e., the phase distribution is approximately Gaussian). When $\ell_D > L$, or the upper-right of Fig. \ref{fig: param map}a, the motional averaging regime emerges where the phase is again approximately Gaussian, but this time via a central limit theorem argument \cite{Neuman1974, Hurlimann1995}. Lastly, note that the diffusion-weighting $b$-value for the CGSE \cite{Carr1954, Stejskal1965, LeBihan1986}, given by $b = (2/3)\gamma^2g^2T^3$, is equivalently expressed as
\begin{equation}\label{eq: bD equiv}
    bD = \frac{2}{3}\left(\frac{\ell_D}{\ell_g}\right)^6
\end{equation}
Thus, when the ratio $\ell_D/\ell_g > 2$, or the bottom part of Fig. \ref{fig: param map}a, one has that $bD \gtrsim 40$ and the signal should be nearly fully dephased such that any contrast with $t_m$ would be difficult to detect, even if there were strong localization. This leaves just the described region. See also figure 2 in ref. \cite{Hurlimann1995} and figure 1 in ref. \cite{Sen1999} for analogous non-dimensional plots. Note that according to these works, the region identified here would be intermediate, i.e., lying between characteristic regimes. This is because what is usually considered to be the localization regime is strongly dephased, whereas here we also take into account the visibility of exchange contrast, $\beta_1 \ge 0.02$. 

In Fig. \ref{fig: param map}b, the corresponding values of $k$ from Fig. \ref{fig: param map}a are plotted on the same axes, filtering for $\beta_1 \ge 0.02$. One sees that the values of $k$ are highly homogeneous (see the color bar), and are all $\approx 50\;\mathrm{s^{-1}}$. This is a preliminary indication that $k$ due to localization has weak dependence on $g$ and $T$, which were the parameters varied to generate the fits. The resolution is coarse, however, and there appears to be a trend of increasing $k$ toward smaller ratios that is worth investigating. 

In Fig. \ref{fig: param map}c, the region of interest $\ell_D/2 < \ell_g < \ell_D < L$ is looked at with finer resolution. Signals were generated for $\ell_D$ and $\ell_g$ spaced linearly by $0.01L = 0.2\;\mathrm{\mu m}$, from $0.01L$ up to $L$ and $0.6L$, respectively, keeping other parameters the same. Values of $k$ are shown, again filtering for $\beta_1 \ge 0.02$. Note the change in axes and color scale. One sees that the majority of the region is homogeneous with $k\approx 50\;\mathrm{s^{-1}}$, but as both $\ell_D/L$ and $\ell_g/L$ become small $\lesssim 0.2$, $k$ increases to order $\sim 100\;\mathrm{s^{-1}}$. Nonetheless, the statement that $k$ has weak dependence on $g$ and $T$ is broadly valid outside of this special case/region. Note that the fit in Fig. \ref{fig: dexsy fit g = 0.3} corresponds to $\ell_D/L \approx 0.22$ and $\ell_g/L \approx 0.07$, which lies near the bottom left of Fig. \ref{fig: param map}c, although the exact point is not shown. In that fit, we observed $k \approx 64\;\mathrm{s^{-1}}$. We return to these results later and provide a physical explanation for the relative homogeneity of $k$ and the exception for small $\ell_D/L$ and $\ell_g/L$. 

\subsection{Exchange rate dependencies}

\begin{figure}
    \centering
    \includegraphics{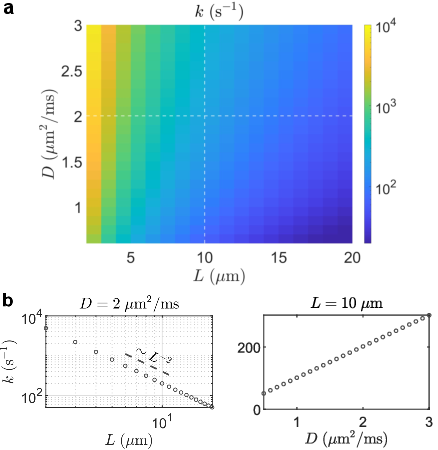}
    \caption{Fitted $k$ for fixed ratios of $\ell_D/L = 0.5$ and $\ell_g/L = 0.3$, and $L$ spaced linearly by $1$ from $2-20\;\mathrm{\mu m}$ and $D$ by $0.1$ from $0.5 - 3\;\mathrm{\mu m^2/ms}$. Discretization was $\Delta t = 2\;\mathrm{\mu s}$ with $\Delta x$ adjusted to maintain $p \approx 0.1$. (a) Values of $k$ on a log color scale. Dashed lines indicate 1-D cross-sections shown in the next part. (b) 1-D cross sections of (a) at fixed $D = 2\;\mathrm{\mu m^2/ms}$ (left) and $L = 10\;\mathrm{\mu m}$ (right). The plot vs. $L$ is on a log-log axis and the $L^{-2}$ dependence is illustrated.}
    \label{fig: k vs L D}
\end{figure}

 We next explore whether and how $k$ depends on $L$ and $D$. Let us isolate an optimal set of lengthscales: $\ell_D/L = 0.5$ and $\ell_g/L = 0.3$, which lies near the maximum $\beta_1$ shown in Fig. \ref{fig: param map}a. Signals were generated for $D$ linearly spaced by $0.1$ from $0.5 - 3\;\mathrm{\mu m^2/ms}$ and $L$ linearly spaced by $1$ from $2 - 20\;\mathrm{\mu m}$. Values of $g$ and $T$ were adjusted to keep the ratios specified above, with $T$ rounded to the nearest tenth of a millisecond. For discretization, $\Delta t = 2\;\mathrm{\mu s}$, while $\Delta x$ was adjusted with $D$ to maintain $p \approx 0.1$, i.e., $\Delta x = \sqrt{10D\Delta t}$ from Eq. \eqref{eq: p}, rounded to the nearest tenth of a micron. In Fig. \ref{fig: k vs L D}a, the $k$ values from fits of Eq. \eqref{eq: 3 param fit} are shown on a log color scale. In Fig. \ref{fig: k vs L D}b, cross-sections at fixed $D = 2\;\mathrm{\mu m^2/ms}$ or $L = 10\;\mathrm{\mu m}$ are shown. The plot with $L$ is shown with log-log axes. These plots reveal approximate $L^{-2}$ and $D$ proportionalities for $k$ such that $k$ becomes very large $\gtrsim 10^3\;\mathrm{s^{-1}}$ in the upper left corner of Fig. \ref{fig: k vs L D}a, where $L \lesssim 5\;\mathrm{\mu m}$ and $D\gtrsim 2\;\mathrm{\mu m^2/ms}$. 

What is the origin of these proportionalities? Consider that any arbitrary $\mathbf{m}(T)$ can be represented in terms of the eigenmodes of the diffusion operator. For reflecting boundaries, these are cosine modes with (non-zero) eigenvalues:
\begin{equation}
\lambda_n = D \left(\frac{\pi n}{L}\right)^2, \quad n > 0.
\end{equation}
As the equilibration process with $t_m$ is diffusive, the decay of these modes should proceed with rates $\lambda_n$. It then follows that $k$, irrespective of the detailed decomposition of $\mathbf{m}(T)$, will have approximate proportionality
\begin{equation}
    k \sim D/L^2.
\end{equation}
Put simply, within the region highlighted in Fig. \ref{fig: param map}, DEXSY is sensitive to the relaxation of spatial magnetization modes that are established by the first CGSE. 

Although this is not a first-order rate process in a physical sense, it can be approximately modeled as such when the first (non-zero) mode dominates. This is somewhat the case seen in Figs. \ref{fig: profiles vs g} and \ref{fig: dexsy fit g = 0.3} --- the profile is roughly a (shifted and scaled) cosine of the form $\cos(\pi x/L)$. We point out that for $L = 20\;\mathrm{\mu m}$ and $D = 2\;\mathrm{\mu m^2/ms}$, $\lambda_1 = \pi^2 D/L^2 \approx 49\;\mathrm{s^{-1}}$, in agreement with $k$ in the homogeneous regions of Figs. \ref{fig: param map}b -- c. Thus we can refine the proportionality to an approximation in this region:
\begin{equation}
    k \approx \frac{\pi^2D}{L^2}, \quad \frac{\ell_D}{L}, \frac{\ell_g}{L} \gtrsim 0.2
\end{equation}
When higher modes are needed to represent $\mathbf{m}(T)$, the signal decay $S(t_m)$ should become multi-exponential, though the first mode should still dominate the longer $t_m$ behavior. Note that the slight biases seen in Fig. \ref{fig: dexsy fit g = 0.3} at short times may reflect such higher mode contributions.

That being said, with this view in mind, the fact that $k$ has weak dependence on $g$ and $T$ is no longer surprising --- these parameters affect $k$ only through their effect on the shape of $\mathbf{m}(T)$ and its resulting decomposition. If these shapes tend to be similar, then said effect is weak. This line of reasoning provides an avenue by which to explore the physical origin of the trend seen in Fig. \ref{fig: param map}c, where $k$ increases as the ratios $\ell_D/L$ and $\ell_g/L$ become small. That is, do the profiles with small ratios truly have more higher-mode content than those with larger ratios, resulting in larger $k$? 

Let us show examples of eigen-decomposition of $\mathbf{m}(T)$. Specifically, we consider the eigenvectors of the discrete, finite difference diffusion operator $\mathbf{W}_{\text{FD}} = (\mathbf{A} - \mathbf{I})/\Delta t$, and denote them as column vectors $\mathbf{u}_n$ with eigenvalues $\lambda_n$. These should approximate the cosine modes consistent with the boundary problem in continuous space. The eigenvectors are taken to be orthonormal, i.e., $\mathbf{u}_n^{\text{T}}\mathbf{u}_n = 1$. Then,
\begin{equation}
    \mathbf{m}(T) \approx \sum_{n=0}^{N_x - 1} \mathbf{u}_n c_n, \quad c_n = \mathbf{u}_n^{\text{T}}\, \mathbf{m}(T),
\end{equation}
where the coefficients $c_n$ are given by projection of $\mathbf{m}(T)$ onto the eigenspace. The magnetization after $t_m$ (before the second encoding) can be approximated as
\begin{equation}
    \mathbf{m}(T + t_m) \approx \sum_{n =0}^{N_x - 1} \mathbf{u}_n c_n \exp(-\lambda_n t_m).
\end{equation}
In this way, the rate of equilibration and thereby signal decay with $t_m$ is tied to the spectrum of $\mathbf{m}(T)$.

\begin{figure}
    \centering
    \includegraphics{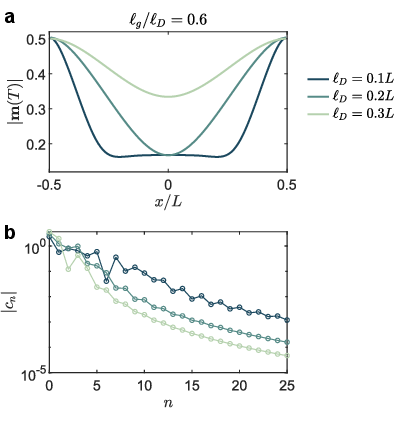}
    \caption{Examples of eigen-decomposition of $\mathbf{m}(T)$. The ratio $\ell_g/\ell_D = 0.6$ was fixed, while $\ell_D = [0.1,\,0.2,\,0.3]L$ was varied (light to dark, respectively). (a) Absolute profiles $|\mathbf{m}(T)|$. Note that fixing $\ell_g/\ell_D$ yields profiles with similar maxima as $bD \propto (\ell_D/\ell_g)^6$. (b) Absolute eigen-decomposition coefficients $c_n$ in the basis $\mathbf{u}_n$, plotted on a log $y$-axis. Note the difference in tails, with increasing high-frequency content as $\ell_D$ and $\ell_g$ decrease. The corresponding values of $k$ (fits not shown) are $\approx [310,\,130,\,53]\;\mathrm{s^{-1}}$, in legend order.}
    \label{fig: mT decomp}
\end{figure}

In Fig \ref{fig: mT decomp}a, profiles $|\mathbf{m}(T)|$ are shown for a fixed ratio of $\ell_g/\ell_D = 0.6$, while varying $\ell_D = [0.1,\,0.2,\,0.3]L$. All other parameters are kept the same as Fig. \ref{fig: param map}. Note that these values of $\ell_D,\, \ell_g$ are chosen to move diagonally through the region-of-interest identified in Fig. \ref{fig: param map}, starting in the area with elevated $k$ (see Fig. \ref{fig: param map}c) and moving into the more homogeneous area when $\ell_D/L = 0.3$. In Fig. \ref{fig: mT decomp}b, the absolute coefficients $|c_n|$ are shown up to mode $n = 25$. One sees that indeed the smallest $\ell_D/L = 0.1$ has more higher mode content. See in particular the tails of the spectra. The corresponding values of $k$ (fits not shown), are $k\approx [310,\,130,\,53]\;\mathrm{s^{-1}}$, respectively.

One aspect of strong localization is that the persistent or more coherent boundary layer of signal scales in relative size as $\ell_g/L$ --- see Stoller \textit{et al.} \cite{Stoller1991}. Thus, as this ratio becomes small, a plateau in $\mathbf{m}(T)$ emerges in the center of the domain, as seen clearly in Fig. \ref{fig: mT decomp}a. It is this plateaued shape that requires higher spatial frequencies represent, driving increased $k$. When localization is weaker (or $\ell_g/L$ larger), the shape is roughly uni-periodic on the domain (see again Figs. \ref{fig: profiles vs g} and \ref{fig: mT decomp}a), hence there is less higher mode content as confirmed by Fig. \ref{fig: mT decomp}b, and the decay of $S(t_m)$ should be closer to monoexponential.   

Though this spectral interpretation is elegant, we caution against trying to predict $k$ via $c_n$ (e.g., as a weighted average of $\lambda_n$), as $k$ is observed only via a phenomenological fit. The fit may obfuscate the spectrum and blur any higher-frequency content. In other words, $k$ is a phenomenological projection of a multiexponential process. For now, we can say only that small $\ell_D/L$ and $\ell_g/L$ leads to larger $k$, driven by spectral features of $\mathbf{m}(T)$. It should also be noted that small ratios are difficult to achieve in practice, requiring large $g$ and short $T$ unless $L$ is extremely large. Consider that the values used in Fig. \ref{fig: mT decomp} are $g \approx [4.3,\,0.55,\,0.16]\;\mathrm{T/m}$ and $T = [2,\,8,\,18]\;\mathrm{ms}$, respectively, where the former $g > 4 \;\mathrm{T/m}$ and $T \lesssim 2 \;\mathrm{ms}$ can be achieved at present only on stray field or static gradient hardware \cite{Casanova2011}. In the vast majority of experimental cases, one is likely to be in the homogeneous region where $k \approx \pi^2D/L^2$.

\section{Discussion}

\subsection{Summary of findings}

We have investigated whether a realization of DEXSY with identical CGSE encodings, characterized by gradient amplitude $g$ and encoding time $T$, can yield apparent exchange contrast with respect to mixing time $t_m$ in a single, one-dimensional compartment of length $L$ with reflecting boundaries. To aid our investigation, we presented an approach to generate signals based on a first-order, Lie-Trotter splitting of the Bloch-Torrey evolution, which was then validated against MC simulations. Using this approach, we identified a parameter range or region in which apparent exchange can be detected, consistent with the localization regime. This serves as a minimal counterexample to the assertion that DEXSY is uniquely sensitive to exchange between distinct compartments or geometric domains. 

In terms of the characteristic lengthscales $\ell_D = \sqrt{DT}$ and $\ell_g = (D/\gamma g)^{1/3}$, this region can be described as $\ell_D/2 < \ell_g < \ell_D < L$. In most of this region, the associated rate constant $k$, which is obtained via a phenomenological three-parameter fit, is approximately $\pi^2 D/L^2$, and is more generally $\sim D/L^2$. This value is equivalent to the first non-zero eigenvalue of the diffusion operator. We interpret localization-driven exchange as being related to the relaxation of spatial magnetization modes established by the first CGSE with $t_m$. When the ratios $\ell_D/L$ and $\ell_g/L$ are both small $\lesssim 0.2$, it was noted that $k$ is increased. This is due to the profile at the first CGSE having a plateaued region in the center of the domain when said ratios are small, which requires higher modes to represent. 

\subsection{Implications in realistic samples}

What do these results imply for DEXSY studies of realistic samples such as neural tissue? We first point out that the region $\ell_D/2 < \ell_g < \ell_D < L$ is not atypical in terms of experimental and physical parameters. Recalling Eq. \eqref{eq: bD equiv}, it corresponds to $1 \lesssim bD \lesssim 40$ and $\ell_D < L$. This translates to whenever there is substantial signal decay and the domain is also larger than the typical root-mean-square displacement. Importantly, we note that the former condition $1 \lesssim bD \lesssim 40$ is not constrained by gradient hardware \textit{per se}, as it is the ratio $\ell_D/\ell_g \propto bD$ that is relevant. In principle, any gradient amplitude can achieve large $b$ with sufficient encoding time $T$, if $R_2$ relaxation permits.  

Let us provide some concrete values. For $D = 3\;\mathrm{\mu m^2/ms}$ and $T$ from $10 - 100\;\mathrm{ms}$, encompassing typical encoding times in conventional scanners, one has $\ell_D = 5.5 - 17\;\mathrm{\mu m}$. Given that soma have diameters that range from $\sim 5 - 20\;\mathrm{\mu m}$ \cite{AirdRossiter2026}, one may encounter structures larger than $\ell_D$ in the study of biological tissue, particularly for short encodings. As somata comprise between $\sim 10 - 40\%$ of gray matter by volume \cite{Keller2018, Ianus2022, ShapsonCoe2024}, the possibility of localization effects in a fraction of neural tissue cannot be neglected. Of course, a similar assessment could be performed for other tissue types and components, or really any sample and/or parameter range, as the region is expressed non-dimensionally.

How exactly localization would impact the measurement of $k$ in biological samples is not obvious, however. Here we have considered a minimal system, isolating localization to yield a clean spectral result. A more realistic system would include relaxation mechanisms, higher dimensionality, a continuum of lengthscales, membrane permeability, geometric heterogeneity, etc. All of these factors may act in tandem to affect $k$. 

Some of these factors are not expected to affect the result. Uniform relaxation mechanisms, namely $R_1$ relaxation during $t_m$ and $R_2$ relaxation during the encodings, would scale all modes equally and would not affect the spectral interpretation of $k$ if properly normalized. Regarding higher dimensions, we point out that the eigenvalue scaling is conserved such that the core result of $k$ scaling with $D$ over the squared structural length is likewise conserved, though the prefactor(s) will depend on the geometry under consideration.

Other factors may affect the result. Surface relaxation, for instance, modifies the boundary condition from Neumann to Robin form, $D \partial_x m + \rho m =0$, where $\rho$ is the surface sink strength density \cite{Brownstein1979}. This introduces another lengthscale, $\ell_\rho = D/\rho$, which can be thought of as the distance over which diffusion can replenish magnetization lost at the surface. When $\ell_\rho \gg L$, surface effects are weak and the results described here should hold. At intermediate values $\ell_\rho \lesssim L$, surface relaxation may compete with gradient-induced localization, particularly when one has the ordering: $\ell_\rho \lesssim \ell_g < \ell_D < L$. This may reduce spatial heterogeneity in the magnetization profile and thereby localization-driven contrast. Interestingly, we note that when $\ell_\rho \ll L$ (i.e., the ``slow-diffusion'' regime of Brownstein and Tarr \cite{Brownstein1979}), surface relaxation may become strong enough to ``invert'' the magnetization profile, i.e., the profile may have a central peak \cite{Afrough2024} instead of a trough as seen here. In this case, intra-compartment exchange may proceed in a similar manner with $t_m$, though via different spectra in an eigenbasis consistent with the Robin boundary conditions.

Geometric heterogeneity and/or membrane permeation may act in a similar way to surface relaxation. That is, they should both diminish the localization effect. In the former, consider that diffusion along an orthogonal branch provides a pathway by which localized signal can escape and dephase. Similarly for the latter, signal can escape through the barrier (e.g., see the magnetization profiles in figure 4 of ref. \cite{Grebenkov2014}). For branching domains, we speculate that the exchange processes should not interact, i.e., localized signal is just as likely to diffuse along a branch vs. non-localized signal. For permeation, however, there is an interaction as localization biases the signal to be near barriers. This may amplify the influence of permeability on signal evolution \cite{Grebenkov2014}. If $k$ is interpreted to be proportional to the surface-to-volume ratio (SVR) in the sense that $k = \kappa \times \text{SVR}$ in the barrier-limited case, where $\kappa$ is permeability, localization may inflate the effective SVR and thus the measured $k$ for the barrier. In this way, $k$ from barrier permeation may depend on $g$, which is an avenue of future study.

Another point to consider is heterogeneous lengthscales, i.e., if $L$ is distributed. Even if domains are non-communicating in that they cannot exchange magnetization, the $k$ that is observed will be influenced by all domains, arising from a complicated superposition of spectra. We point out that small domains will have much faster $k$ due to the $L^{-2}$ scaling --- e.g., for $L = 10\;\mathrm{\mu m}$ and $D = 3\;\mathrm{\mu m^2/ms}$, one has $\pi^2D/L^2 \approx 300\;\mathrm{s^{-1}}$ which would equilibrate over just a few milliseconds. Larger domains may dominate the longer-$t_m$ signal behavior and have more influence on the fit. We also note that the amount of contrast $\beta_1$ varies (see again Fig. \ref{fig: param map}a) with a maxima around $\ell_D/L \approx 0.5$, $\ell_g/D \approx 0.35$. This is another source of bias that emphasizes domains near this maxima for the given $\ell_D$ and $\ell_g$. Thus, we suspect that the observable $k$ will be biased towards certain domains, rather than being proportionally weighted by volume or density. In addition, there may be heterogeneous $R_2$ and surface relaxivity, which can also produce exchange contrast \cite{Lee1993, Washburn2006, Ordinola2024}.  

In summary, even if localization-driven exchange is expected in terms of the characteristic lengthscales, the interpretation that $k\approx \pi^2 D/L^2$ may be complicated or even confounded by myriad factors: surface relaxivity, alternative exchange mechanisms, and lengthscale heterogeneity. The former two can decrease the localization effect. All of these may contribute to highly multi-exponential behavior in the signal. Thus, the localization effect described here should be regarded as one \emph{possible} source of DEXSY signal contrast, not necessarily as a dominant or even likely one. Note that the maximal contrast from this mechanism is $\sim 0.1$ (in terms of normalized signal, see again Fig. \ref{fig: param map}a) such that if there is greater contrast with $t_m$, there are likely other processes in play. The only case in which the interpretation here is exact is when the sample is a closed pore or homogeneous collection of pores with non-relaxing boundaries. Again, the primary implication of this work is that DEXSY is not specific to membrane permeation.

\subsection{Extensions and outlook}

Including and accounting for the above effects is a natural extension. Including more reflecting boundaries (i.e., partitioning the domain) is trivial. Uniform relaxation mechanisms can be accounted for by interleaving an additional diagonal matrix operator with entries of the form $\exp(-R\Delta t)$. To include higher dimensions, one can add shifted diagonals to $\mathbf{A}$ that represent motion along different spatial axes, while also flattening $\mathbf{m}$ and extending the diagonal operator matrices --- e.g., see ref. \cite{Cai2025} for a two-dimensional implementation of $\mathbf{A}$. Even advective drift can be included by making $\mathbf{A}$ asymmetric in accordance to an Eulerian specification of the flow field.  

Regarding boundary conditions, a relaxing boundary can be modeled by modifying the first and last entries in $\mathbf{A}$ from $1 - p$ to $1 - p - \rho(\Delta t/\Delta x)$ \cite{Sen1999}. As an extension, imagine a third non-dimensional axis $\ell_\rho/L$ in Fig. \ref{fig: param map}, along which we speculate that exchange contrast may vanish around $\ell_\rho/L \sim 1$, but persists with (potentially different) rates at both extremes, as discussed. Permeable barriers can be modeled by modifying the elements in $\mathbf{A}$ adjacent to such a barrier from $p$ and $1 - 2p$ to $pp_\kappa$ and $1 - p - pp_\kappa$, where $p_\kappa$ represents a transition probability related to the physical permeability by $p_\kappa = \kappa \Delta x/(D + \kappa \Delta x)$ \cite{Cai2025, Herberthson2025}. These boundary conditions can also be combined as the term $\rho(\Delta t/\Delta x)$ is additive. The signal generation framework is thus highly flexible and can be adjusted to explore the speculations above, such as whether can surface relaxivity drive exchange contrast when $\ell_\rho \ll L$, how exactly localization and barrier permeation interact, etc. 

That said, let us consider the case where the interpretation $k \approx \pi^2 D/L^2$ holds --- closed, non-relaxing pore(s). In this case, localization-driven exchange is not merely an artifact, but could serve as a means to estimate the pore size. Consider that large pores can be difficult to probe with conventional PGSE as the deviation from free diffusion may be small $L \gg \ell_D$ until the signal is strongly dephased, which may require long encodings leading to $R_2$ relaxation issues. The experiment here may be a preferable alternative as it uses a longitudinal storage period $t_m$ (typically, $R_1 < R_2$). For a proof-of-principle on leveraging localization to estimate size, see ref. \cite{Lee2023}.

In a similar vein, the localization effect in the first CGSE can be seen as an excitation of higher eigenmodes which then relax during $t_m$ and are observed or read out during the second CGSE. There is potentially rich information contained in this signal behavior, which we only observe in a crude sense via a monoexponential fit. A more involved analysis approach such as a numerical inverse Laplace transform has the potential to reveal the excited spectrum as shown in Fig. \ref{fig: mT decomp}b. This may be a useful way to probe high eigenmodes and thereby information about the shape of the domain \cite{Kac1966}. Parallels may be drawn to methods such as diffusion pore imaging \cite{Laun2011} and ref. \cite{Song2000} which uses inhomogeneous internal magnetic fields. We also show in \ref{appx: FEXSY} that the second CGSE which functions as a readout does not need to be identical to the first, broadening the design space of the method. Moreover, the first encoding may be designed to vary how it excites the spectrum. For instance, an oscillating gradient spin echo (OGSE) \cite{Callaghan1995, Parsons2005} might excite high modes selectively.   

\section*{Conclusion}

Though DEXSY and related methods are commonly believed to be sensitive to barrier permeation, there is growing evidence that other phenomena can also produce exchange-like signal contrast. We detail one such phenomenon that was previously unstudied, in which signal localization sensitizes DEXSY to the relaxation of spatial magnetization modes. Clear criteria for when this phenomenon may be expected were provided in terms of non-dimensional lengthscale ratios. Additionally, we showed that for a reflecting, one-dimensional compartment, the observed exchange rate is typically consistent with the first non-zero eigenvalue of the Laplacian basis. Our results highlight the potential complexity of DEXSY and double diffusion encoding signals, even in seemingly simple systems.

\appendix
\setcounter{figure}{0}
\section{Full DEXSY sampling with numerical inverse Laplace transform}

In the main text, we considered only a realization of DEXSY in which the encodings are identical. There are a variety of ways to sample and analyze DEXSY data in the literature \cite{Ordinola2023}. We consider two other approaches in these Appendices. In its original conception, DEXSY involves a numerical, two-dimensional inverse Laplace transform (ILT) \cite{Callaghan2004}, where the variables are the $b$-value in each encoding, denoted $b_1$ and $b_2$. One assumes continuously distributed Gaussian compartments such that the signal is described by
\begin{equation}
    S(t_m) = \int_0^\infty \int_0^\infty \exp{\left(-b_1D_1 -b_2D_2\right)}P(D_1,D_2,t_m)\, dD_1 dD_2,
\end{equation}
where $P(D_1, D_2, t_m)$ is the joint probability density function (PDF) of the diffusivities during each encoding, $D_1$ and $D_2$. Off-diagonal content in the PDF as recovered by an ILT is thought to be indicative of exchange. 

Signals were generated by the method in the main text with parameters $L = 20\;\mathrm{\mu m}$, $T = 30\;\mathrm{ms}$, $D = 2\;\mathrm{\mu m^2/ms}$, $\Delta x = 0.2\;\mathrm{\mu m}$, $\Delta t = 2\;\mathrm{\mu s}$, and $p = 0.1$. This gives $\ell_D/L \approx 0.39$ and thus we expect to be in the region where $k \approx \pi^2D/L^2\approx 50\;\mathrm{s}^{-1}$ --- see again Fig. \ref{fig: param map}c. Values of $g$ in each encoding were chosen to yield a $40\times 40$ $b$-value grid, uniformly and linearly spaced between $0 - 40\;\mathrm{ms/\mu m^2}$, i.e., $bD$ ranges from $0 - 80$. 

For the ILT, we implemented a non-negative least squares optimization with regularization, detailed below. Let $K(b, D) = \exp(-bD)$ denote the signal kernel per encoding, with discrete form $\mathbf{K}_{ij}$ for the $i^{\text{th}}$ $b$-value and $j^{\text{th}}$ diffusivity, and we note that the $b$-values are uniform such that the same $\mathbf{K}$ may be used for both encodings. We assume a grid of $60$ values of $D$ log-linearly spaced between $10^{-3} - 10^1\;\mathrm{\mu m^2/ms}$. A solution for the discrete PDF, denoted $\mathbf{P}$, can be found by minimizing
\begin{equation}
    || \mathbf{K} \mathbf{P}\mathbf{K}^{\text{T}} - \mathbf{S}||^2 + \lambda || \mathbf{P} ||^2,
\end{equation}
where $||\cdot ||$ denotes the norm, $\mathbf{S}$ is the signal on the $b$-value grid for some $t_m$, and $\lambda$ here denotes a regularization parameter. We take a gradient descent approach wherein $\mathbf{P}$ is updated as
\begin{equation}
    \mathbf{P} \leftarrow \mathbf{P} - \alpha \left[\mathbf{K}^{\text{T}}(\mathbf{K} \mathbf{P}\mathbf{K}^{\text{T}} - \mathbf{S})\mathbf{K} + \lambda \mathbf{P}\right],
\end{equation}
where $\alpha$ is an update velocity set to $\alpha = (|| \mathbf{K} ||^4 + \lambda)^{-1}$. Also, $\mathbf{P}$ is constrained at each update to be non-negative, with negative entries set to $0$. Gaussian white noise was added to $\mathbf{S}$ at an SNR of 100 prior to the ILT. The regularization parameter was set to $\lambda = 10^{-4}$ based on a rough L-curve estimate \cite{Hansen2000} (data not shown). Lastly, $\mathbf{P}$ was normalized to numerically integrate to $1$. 

\begin{figure*}
    \centering
    \includegraphics{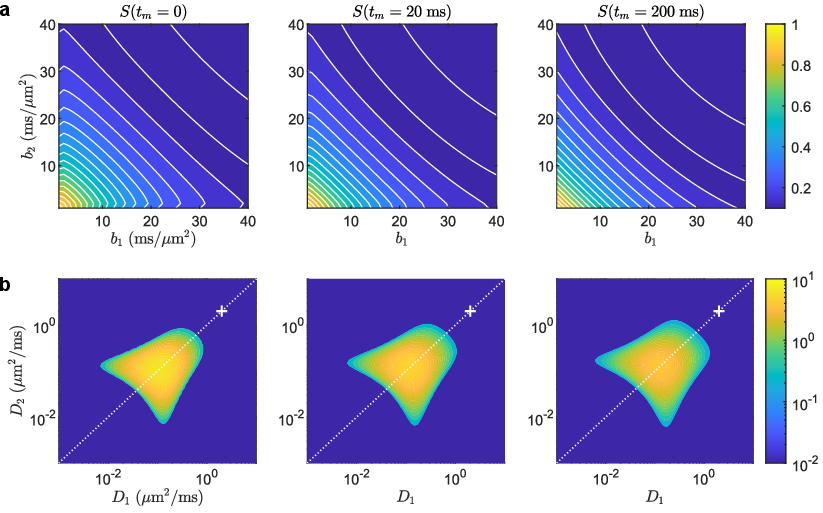}
    \caption{Signals and spectra $P(D_1, D_2, t_m)$ from numerical ILT. Signals were generated with $L = 20\;\mathrm{\mu m}$, $T = 30\;\mathrm{ms}$, $D = 2\;\mathrm{\mu m^2/ms}$, $\Delta x = 0.2\;\mathrm{\mu m}$, $\Delta t = 2\;\mathrm{\mu s}$, and $p = 0.1$, with $g$ varied to produce a uniform grid of $40\times 40$ $b$-values linearly spaced from $0 - 40\;\mathrm{ms/\mu m^2}$. Three values of $t_m = [0, 20, 200]\;\mathrm{ms}$ were considered (left-to-right). (a) Contour maps of the signal, with isolines linearly spaced from $0 - 1$ by $0.05$. Note the off-diagonal curvature as $t_m$ increases. (b) Corresponding spectra $P(D_1, D_2)$ from a regularized, non-negative least squares implementation of the ILT, shown with a log color scale. These are normalized to numerically integrate to $1$. For discretization, 60 values of $D_1$, $D_2$ were used, log-linearly spaced between $10^{-3} - 10^{1}\;\mathrm{\mu m^2/ms}$. Gaussian noise was added to the signal at an SNR of 100 before the ILT. The line $D_1 = D_2$ is shown (dotted), as is the intrinsic diffusivity (cross). Note the single, diffuse peak and its spreading behavior orthogonal to $D_1 = D_2$ as $t_m$ increases. }
    \label{fig: ILT}
\end{figure*}

In Fig. \ref{fig: ILT}b, a contour map of $\mathbf{S}$ at $t_m = [0,\,20,\,200]\;\mathrm{ms}$ (left-to-right) is shown, with the contours linearly spaced from $0 - 1$ by $0.05$. The value $t_m = 20\;\mathrm{ms}$ is about $1/k$. Note that the signal decays very slowly with $b$, considering that we go up to $40\;\mathrm{ms/\mu m^2}$ or $bD = 80$. One can also see the signal contours develop curvature along the forward diagonal for which $b_1 + b_2$ is constant. This is in line with previous work \cite{Cai2018, Cai2022, Song2016} that showed such curvature is indicative of an exchange process. In Fig. \ref{fig: ILT}b, the corresponding spectra, or PDFs $P(D_1, D_2)$ are shown. There is a single, highly diffuse peak, that spreads orthogonally away from the equality line $D_1 = D_2$ (dotted) as $t_m$ increases. The intrinsic diffusivity $D = 2\;\mathrm{\mu m^2/ms}$ is also marked by a cross for comparison.  

Localization-driven exchange does not manifest as separated spectral peaks. This is because the signal kernel $K(b,D) = \exp(-bD)$ fails to describe non-Gaussian signal decay. Decay in the localization regime is, to a first approximation, described instead by $\ln S \propto -(bD)^{1/3}$ \cite{Stoller1991, Moutal2019}. The diffuse peak seen in Fig. \ref{fig: ILT} can be seen as the ILT attempting to approximate a stretched exponential shape using many exponentials, which leads to a broad diffusivity spectrum. See also figure 2 in ref. \cite{Cai2022} for a similar study of diffusion-diffusion spectra under non-Gaussian signal decay. Thus, the ILT approach is misleading in the presence of localization, as it yields a broad spectrum to describe what is actually a single compartment.

Still, it remains true that localization-driven exchange results in greater off-diagonal content in the spectrum (i.e., away from $D_1 = D_2$). The precise way in which the spectrum spreads --- orthogonally from $D_1 = D_2$, or equivalently along the axis $\Delta D = D_1 - D_2$ --- can be thought of as follows. Consider the variance in the (spectral) diffusivity difference $\Delta D$:
\begin{equation}
    \text{Var}\left(\Delta D\right) = \text{Var}\left(D_1\right) + \text{Var}\left(D_2\right) - \text{Cov}\left(D_1, D_2\right).
\end{equation}
The covariance term will decrease with $t_m$ irrespective of the particular or true representation of the signal, simply because the encodings are increasingly decorrelated via time separation. Thus, the variance along the axis $\Delta D$ is expected to increase, and this is what is seen in Fig. \ref{fig: ILT}b from left to right.

\section{Filter exchange spectroscopy (FEXSY)}\label{appx: FEXSY}

FEXSY or FEXI is another approach which views exchange as a recovery process in the ADC. Using the nomenclature of the previous appendix, FEXSY applies a fixed ``filter'' of $b_1$, usually denoted $b_f$ \cite{Aslund2009}, after which $b_2$ is varied to measure an ADC, which we denote as $D_{\text{app}}(t_m)$. As a minimal example, at a given $t_m$,   
\begin{equation}\label{eq: Dapp}
    D_{\text{app}} = \frac{\ln \left(S/ S'\right)}{b_2' - b_2},
\end{equation}
where $b_2$ and $b'_2$ denote two $b$-values in the second encoding. One then observes the recovery of $D_{\text{app}}(t_m)$ to extract $k$:
\begin{equation}\label{eq: Dapp fit}
    D_{\text{app}}(t_m) = D_{\infty} + (D_0 - D_{\infty})\exp{\left(-kt_m\right)},
\end{equation}
where $D_\infty$ is $\lim_{t_m\rightarrow \infty} D_{\text{app}}(t_m)$ and $D_0$ is an apparent intrinsic diffusivity, not necessarily equal to $D$. Although this is written in terms of ADC values, it is functionally identical to Eq. \eqref{eq: 3 param fit} in the sense that one has a limiting value $D_{\infty}$ (vs. $\beta_3$), a total variation $D_0 - D_{\infty}$ (vs. $\beta_1$), and an exponential rate constant $k$. 

In fact, the experiment in the main text can be viewed as a special case of FEXSY where $b_1$ or $b_f = b_2$ and $b_2' = 0$, wherein instead of extracting an ADC, we leave the fit in terms of the raw signal --- see again Eqs. \eqref{eq: Dapp} -- \eqref{eq: Dapp fit} above and compare to Eq. \eqref{eq: 3 param fit}. Note that Eq. \eqref{eq: Dapp} can also be thought as a finite difference derivative approximation of the log signal, $D_{\text{app}} = \partial_{b_2} \ln S$, which preserves the exponential term if we substitute $S$ in the form of Eq. \eqref{eq: 3 param fit}. This means that we can expect to measure the same $k$ as in the main text when the $b$-values are chosen accordingly. As a corollary, FEXSY should be sensitive to localization when the filtering encoding is in the described regime of $\ell_D/2 < \ell_g < \ell_D < L$. 

That said, we can use this appendix to explore here whether the result $k \approx \pi^2D/L^2$ holds for $b_1 \neq b_2$, as FEXSY makes no explicit prescription for the choice of $b_2$ values. One might, for example, want to use small $b_2$ to preserve signal if $b_1$ or $b_f$ is large. We keep the parameters used in the previous appendix, varying only $g$ in each encoding. We choose $b_f \approx 4.6\;\mathrm{ms/\mu m^2}$, corresponding to $g = 0.06 \;\mathrm{T/m}$, $\ell_g/L\approx 0.25$, and again $\ell_D/L \approx 0.39$. Looking at Fig. \ref{fig: param map}c, we again expect to be in the homogeneous region of $k$ for these parameters. For $b_2$, we choose $\approx [0.13,\,0.52]\;\mathrm{ms/\mu m^2}$ to measure $D_{\text{app}}$, corresponding to $g = [0.01,\, 0.02]\;\mathrm{T/m}$, respectively. For $t_m$, we keep the same sampling as in Fig. \ref{fig: dexsy fit g = 0.3}, with 30 values of $t_m$ log-linearly spaced from $10^{-1}$ -- $10^{2.5}\;\mathrm{ms}$. 

\begin{figure}
    \centering
    \includegraphics{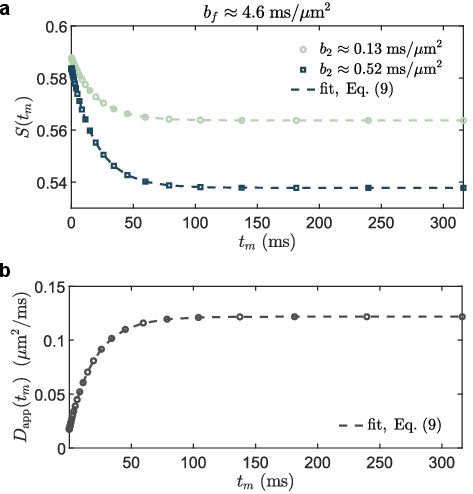}
    \caption{Signals and $D_{\text{app}}(t_m)$ for an example FEXSY experiment. Signal generation parameters were the same as Fig. \ref{fig: ILT}, and $t_m$ were the same as Fig. \ref{fig: dexsy fit g = 0.3}. The filter $b_f \approx 4.6\;\mathrm{ms/\mu m^2}$ corresponds to $g = 0.06\;\mathrm{T/m}$, while $b_2 \approx [0.13,\, 0.52]\;\mathrm{ms/\mu m^2}$ correspond to $g = [0.01, 0.02]\;\mathrm{T/m}$ (light green and dark blue, respectively). (a) Raw signals $S(t_m)$. A fit of Eq. \eqref{eq: 3 param fit} is also shown (dashed line), yielding $k\approx 50\;\mathrm{s^{-1}}$ in both cases. (b) Corresponding $D_{\text{app}}(t_m)$ calculated by Eq. \eqref{eq: Dapp}. A fit of Eq. \eqref{eq: 3 param fit} yields a similar $k\approx 48\;\mathrm{s^{-1}}$. Note that since this is a recovery process, $\beta_1$ in the fit is negative.}
    \label{fig: FEXSY}
\end{figure}

In Fig. \ref{fig: FEXSY}a, the raw signals are plotted for the two values of $b_2 \approx [0.13,\,0.52]\;\mathrm{ms/\mu m^2}$. Fits of Eq. \eqref{eq: 3 param fit} are also shown, and both yield $k \approx 50\;\mathrm{s}^{-1}$, consistent with $\pi^2D/L^2 \approx 49\;\mathrm{s}^{-1}$, as expected. In Fig. \ref{fig: FEXSY}b, the corresponding $D_{\text{app}}$ estimates are shown, calculated using Eq. \eqref{eq: Dapp}. The same fit was used, which yields a similar $k \approx 48\;\mathrm{s}^{-1}$. These results suggest that the second CGSE is not relevant to the behavior with $t_m$, as the same rate is measured at various $b_2$. Thus, we expect that arbitrary readout encodings may be used to probe the same phenomenon. In addition, one sees that interpreting the data in terms of $D_{\text{app}}$ preserves the time-dependence, as confirmed by the similar $k$. This step can be considered ancillary if the goal is to measure $k$, though $D_{\infty}$ and $D_0$ themselves may also be parameters of interest. To conclude, we have verified that our findings in the main text apply to FEXSY. Any $b_f$ encoding consistent with the regime identified should sensitize FEXSY to localization-driven exchange and thereby the diffusion spectrum.

\section*{Contributions}

TXC conceived the research, carried out all simulations and data analysis, and wrote the original draft of the manuscript. PJB supervised the project. All authors reviewed and edited the manuscript. 

\section*{Declaration and data availability}

The authors declare no competing financial interests. The MATLAB code and simulated or generated data are available upon reasonable request.

The views, information or content, and conclusions presented do not necessarily represent the official position or policy of, nor should any official endorsement be inferred on the part of, the Uniformed Services University, the Department of War, the U.S. Government, or The Henry M. Jackson Foundation for the Advancement of Military Medicine, Inc. The contributions of the NIH author(s) are considered Works of the United States Government. The findings and conclusions presented in this paper are those of the author(s) and do not necessarily reflect the views of the NIH or the U.S. Department of Health and Human Services.

\section*{Acknowledgments}

TXC, NHW, and PJB were supported by the intramural research program (IRP) of the \textit{Eunice Kennedy Shriver} National Institute of Child Health and Human Development (NICHD). NHW was funded by the Military Traumatic Brain Injury Initiative (MTBI\textsuperscript{2}) through the Uniformed Services University of the Health Sciences (USU), Bethesda, MD (award No. HU0001-24-2-0051).




\bibliographystyle{elsarticle-num}

\small
\bibliography{refs}

@article{Topgaard2017,
  title = {Multidimensional diffusion MRI},
  volume = {275},
  DOI = {10.1016/j.jmr.2016.12.007},
  journal = {Journal of Magnetic Resonance},
  author = {Topgaard,  Daniel},
  year = {2017},
  pages = {98–113}
}

@article{Benjamini2020,
  title = {Multidimensional correlation MRI},
  volume = {33},
  DOI = {10.1002/nbm.4226},
  number = {12},
  journal = {NMR in Biomedicine},
  author = {Benjamini,  Dan and Basser,  Peter J.},
  year = {2020}
}

@article{Henriques2021,
  title = {Double diffusion encoding and applications for biomedical imaging},
  volume = {348},
  DOI = {10.1016/j.jneumeth.2020.108989},
  journal = {Journal of Neuroscience Methods},
  author = {Henriques,  Rafael N. and Palombo,  Marco and Jespersen,  Sune N. and Shemesh,  Noam and Lundell,  Henrik and Ianuş,  Andrada},
  year = {2021},
  pages = {108989}
}

@article{Callaghan2004,
author = {Callaghan, P. T.  and Furó, I.},
title = {Diffusion-diffusion correlation and exchange as a signature for local order and dynamics},
journal = {The Journal of Chemical Physics},
volume = {120},
number = {8},
pages = {4032-4038},
year = {2004},
doi = {10.1063/1.1642604}
}

@article{Qiao2005,
  title = {Diffusion exchange NMR spectroscopic study of dextran exchange through polyelectrolyte multilayer capsules},
  volume = {122},
  DOI = {10.1063/1.1924707},
  number = {21},
  journal = {The Journal of Chemical Physics},
  author = {Qiao,  Y. and Galvosas,  P. and Adalsteinsson,  T. and Sch\"{o}nhoff,  M. and Callaghan,  P. T.},
  year = {2005}
}

@article{Aslund2009,
title = "Filter-exchange {PGSE} {NMR} determination of cell membrane permeability",
journal = "Journal of Magnetic Resonance",
volume = "200",
number = "2",
pages = "291-295",
year = "2009",
author = "I. \r{A}slund and A. Nowacka and M. Nilsson and D. Topgaard"
}

@article{Ramadan2009,
  title = {Diffusion-Exchange Weighted Imaging},
  volume = {3},
  DOI = {10.4137/mri.s3504},
  journal = {Magnetic Resonance Insights},
  author = {Ramadan,  Saadallah},
  year = {2009},
}

@article{Lasic2011,
  title = {Apparent exchange rate mapping with diffusion MRI},
  volume = {66},
  DOI = {10.1002/mrm.22782},
  number = {2},
  journal = {Magnetic Resonance in Medicine},
  author = {Lasič,  Samo and Nilsson,  Markus and L\"{a}tt,  Jimmy and Ståhlberg,  Freddy and Topgaard,  Daniel},
  year = {2011},
  pages = {356–365}
}

@article{Nilsson2013,
  title={Noninvasive mapping of water diffusional exchange in the human brain using filter-exchange imaging},
  author={Nilsson, M. and L{\"a}tt, J. and {van Westen}, D. and Brockstedt, S. and Lasi{\v{c}}, S. and St{\aa}hlberg, F. and Topgaard, D.},
  journal={Magnetic Resonance in Medicine},
  volume={69},
  number={6},
  pages={1572--1580},
  year={2013}
}

@article{Lampinen2016,
  title = {Optimal experimental design for filter exchange imaging: Apparent exchange rate measurements in the healthy brain and in intracranial tumors},
  volume = {77},
  DOI = {10.1002/mrm.26195},
  number = {3},
  journal = {Magnetic Resonance in Medicine},
  author = {Lampinen,  Bj\"{o}rn and Szczepankiewicz,  Filip and van Westen,  Danielle and Englund,  Elisabet and C Sundgren,  Pia and L\"{a}tt,  Jimmy and Ståhlberg,  Freddy and Nilsson,  Markus},
  year = {2016},
  pages = {1104–1114}
}

@article{Benjamini2017,
  title = {Imaging Local Diffusive Dynamics Using Diffusion Exchange Spectroscopy MRI},
  volume = {118},
  DOI = {10.1103/physrevlett.118.158003},
  number = {15},
  journal = {Physical Review Letters},
  author = {Benjamini,  Dan and Komlosh,  Michal E. and Basser,  Peter J.},
  year = {2017}
}

@article{Cai2018,
  title = {Rapid detection of the presence of diffusion exchange},
  volume = {297},
  DOI = {10.1016/j.jmr.2018.10.004},
  journal = {Journal of Magnetic Resonance},
  author = {Cai,  Teddy X. and Benjamini,  Dan and Komlosh,  Michal E. and Basser,  Peter J. and Williamson,  Nathan H.},
  year = {2018},
  pages = {17-22}
}

@article{Williamson2020,
  title = {Real-time measurement of diffusion exchange rate in biological tissue},
  volume = {317},
  DOI = {10.1016/j.jmr.2020.106782},
  journal = {Journal of Magnetic Resonance},
  author = {Williamson,  Nathan H. and Ravin,  Rea and Cai,  Teddy X. and Benjamini,  Dan and Falgairolle,  Melanie and O’Donovan,  Michael J. and Basser,  Peter J.},
  year = {2020},
  pages = {106782}
}

@article{Bai2020,
  title = {Feasibility of filter-exchange imaging (FEXI) in measuring different exchange processes in human brain},
  volume = {219},
  DOI = {10.1016/j.neuroimage.2020.117039},
  journal = {NeuroImage},
  author = {Bai,  Ruiliang and Li,  Zhaoqing and Sun,  Chaoliang and Hsu,  Yi-Cheng and Liang,  Hui and Basser,  Peter},
  year = {2020},
  pages = {117039}
}

@article{BreenNorris2020,
  title = {Measuring diffusion exchange across the cell membrane with DEXSY (Diffusion Exchange Spectroscopy)},
  volume = {84},
  DOI = {10.1002/mrm.28207},
  number = {3},
  journal = {Magnetic Resonance in Medicine},
  author = {Breen‐Norris,  James O. and Siow,  Bernard and Walsh,  Claire and Hipwell,  Ben and Hill,  Ioana and Roberts,  Thomas and Hall,  Matt G. and Lythgoe,  Mark F. and Ianus,  Andrada and Alexander,  Daniel C. and Walker‐Samuel,  Simon},
  year = {2020},
  pages = {1543–1551}
}

@article{Cai2022,
  title = {Disentangling the Effects of Restriction and Exchange With Diffusion Exchange Spectroscopy},
  volume = {10},
  DOI = {10.3389/fphy.2022.805793},
  journal = {Frontiers in Physics},
  author = {Cai,  Teddy X. and Williamson,  Nathan H. and Ravin,  Rea and Basser,  Peter J.},
  year = {2022},
  pages = {805793}
}

@article{Cai2024,
title = {The Diffusion Exchange Ratio (DEXR): A minimal sampling of diffusion exchange spectroscopy to probe exchange, restriction, and time-dependence},
journal = {Journal of Magnetic Resonance},
volume = {366},
pages = {107745},
year = {2024},
doi = {https://doi.org/10.1016/j.jmr.2024.107745},
author = {Teddy X. Cai and Nathan H. Williamson and Rea Ravin and Peter J. Basser}
}

@article{Williamson2023,
  title = {Water exchange rates measure active transport and homeostasis in neural tissue},
  volume = {2},
  DOI = {10.1093/pnasnexus/pgad056},
  number = {3},
  journal = {PNAS Nexus},
  author = {Williamson,  Nathan H and Ravin,  Rea and Cai,  Teddy X and Falgairolle,  Melanie and O’Donovan,  Michael J and Basser,  Peter J},
  year = {2023},
  pages = {pgad056}
}

@article{Li2025,
  title = {Comparison of water exchange measurements between filter‐exchange imaging and diffusion time‐dependent kurtosis imaging in the human brain},
  volume = {93},
  DOI = {10.1002/mrm.30454},
  number = {6},
  journal = {Magnetic Resonance in Medicine},
  publisher = {Wiley},
  author = {Li,  Zhaoqing and Liang,  Chunjing and He,  Qingping and Feiweier,  Thorsten and Hsu,  Yi‐Cheng and Li,  Jianhua and Bai,  Ruiliang},
  year = {2025},
  pages = {2357–2369}
}

@article{Karger1969,
  title = {Zur Bestimmung der Diffusion in einem Zweibereichsystem mit Hilfe von gepulsten Feldgradienten},
  volume = {479},
  DOI = {10.1002/andp.19694790102},
  number = {1–2},
  journal = {Annalen der Physik},
  author = {K\"{a}rger,  J.},
  year = {1969},
  pages = {1–4}
}

@article{Karger1985,
  title = {NMR self-diffusion studies in heterogeneous systems},
  volume = {23},
  DOI = {10.1016/0001-8686(85)80018-x},
  journal = {Advances in Colloid and Interface Science},
  publisher = {Elsevier BV},
  author = {K\"{a}rger,  J\"{o}rg},
  year = {1985},
  pages = {129–148}
}

@article{Jelescu2022,
  title = {Neurite Exchange Imaging (NEXI): A minimal model of diffusion in gray matter with inter-compartment water exchange},
  volume = {256},
  DOI = {10.1016/j.neuroimage.2022.119277},
  journal = {NeuroImage},
  author = {Jelescu,  Ileana O. and de Skowronski,  Alexandre and Geffroy,  Fran\c{c}oise and Palombo,  Marco and Novikov,  Dmitry S.},
  year = {2022},
  pages = {119277}
}

@article{Jensen2023,
  title = {Diffusional kurtosis time dependence and the water exchange rate for the multi‐compartment K\"{a}rger model},
  volume = {91},
  DOI = {10.1002/mrm.29926},
  number = {3},
  journal = {Magnetic Resonance in Medicine},
  author = {Jensen,  Jens H.},
  year = {2023},
  pages = {1122–1135}
}

@article{AirdRossiter2026,
  title = {Decoding gray matter,  large-scale analysis of brain cell morphometry to inform microstructural modeling of diffusion MR signals},
  volume = {9},
  DOI = {10.1038/s42003-025-09353-5},
  number = {1},
  journal = {Communications Biology},
  author = {Aird-Rossiter,  Charlie and Zhang,  Hui and Alexander,  Daniel C. and Jones,  Derek K. and Palombo,  Marco},
  year = {2026}
}

@article{Keller2018,
  title = {Cell Densities in the Mouse Brain: A Systematic Review},
  volume = {12},
  DOI = {10.3389/fnana.2018.00083},
  journal = {Frontiers in Neuroanatomy},
  author = {Keller,  Daniel and Er\"{o},  Csaba and Markram,  Henry},
  year = {2018}
}

@article{Ianus2022,
  title = {Soma and Neurite Density MRI (SANDI) of the in-vivo mouse brain and comparison with the Allen Brain Atlas},
  volume = {254},
  DOI = {10.1016/j.neuroimage.2022.119135},
  journal = {NeuroImage},
  author = {Ianuş,  Andrada and Carvalho,  Joana and Fernandes,  Francisca F. and Cruz,  Renata and Chavarrias,  Cristina and Palombo,  Marco and Shemesh,  Noam},
  year = {2022},
  pages = {119135}
}

@article{ShapsonCoe2024,
  title = {A petavoxel fragment of human cerebral cortex reconstructed at nanoscale resolution},
  volume = {384},
  DOI = {10.1126/science.adk4858},
  number = {6696},
  journal = {Science},
  author = {Shapson-Coe,  Alexander and Januszewski,  Michał and Berger,  Daniel R. and Pope,  Art and Wu,  Yuelong and Blakely,  Tim and Schalek,  Richard L. and Li,  Peter H. and Wang,  Shuohong and Maitin-Shepard,  Jeremy and Karlupia,  Neha and Dorkenwald,  Sven and Sjostedt,  Evelina and Leavitt,  Laramie and Lee,  Dongil and Troidl,  Jakob and Collman,  Forrest and Bailey,  Luke and Fitzmaurice,  Angerica and Kar,  Rohin and Field,  Benjamin and Wu,  Hank and Wagner-Carena,  Julian and Aley,  David and Lau,  Joanna and Lin,  Zudi and Wei,  Donglai and Pfister,  Hanspeter and Peleg,  Adi and Jain,  Viren and Lichtman,  Jeff W.},
  year = {2024},
}

@article{Khateri2022,
  title = {What does FEXI measure?},
  volume = {35},
  DOI = {10.1002/nbm.4804},
  number = {12},
  journal = {NMR in Biomedicine},
  author = {Khateri,  Mohammad and Reisert,  Marco and Sierra,  Alejandra and Tohka,  Jussi and Kiselev,  Valerij G.},
  year = {2022}
}

@misc{Chakwizira2025,
  doi = {10.48550/ARXIV.2504.21537},
  author = {Chakwizira,  Arthur and Şimşek,  Kadir and Szczepankiewicz,  Filip and Palombo,  Marco and Nilsson,  Markus},
  title = {The role of dendritic spines in water exchange measurements with diffusion MRI: Double Diffusion Encoding and free-waveform MRI},
  publisher = {arXiv},
  year = {2025}
}

@misc{Kiselev2026,
  doi = {10.48550/ARXIV.2601.20657},
  author = {Kiselev,  Valerij G. and Li,  Jing-Rebecca},
  title = {What Does FEXI Measure in Neurons?},
  publisher = {arXiv},
  year = {2026}
}

@article{Hyslop1991,
  title = {Effects of restricted diffusion on microscopic NMR imaging},
  volume = {94},
  DOI = {10.1016/0022-2364(91)90136-h},
  number = {3},
  journal = {Journal of Magnetic Resonance},
  author = {Hyslop,  W. Brian and Lauterbur,  Paul C.},
  year = {1991},
  pages = {501–510}
}

@article{Putz1992,
  title = {Edge enhancement by diffusion in microscopic magnetic resonance imaging},
  volume = {97},
  DOI = {10.1016/0022-2364(92)90235-y},
  number = {1},
  journal = {Journal of Magnetic Resonance},
  author = {P\"{u}tz,  B and Barsky,  D and Schulten,  K},
  year = {1992},
  pages = {27–53}
}

@article{Callaghan1993,
  title = {Diffusive Relaxation and Edge Enhancement in NMR Microscopy},
  volume = {101},
  DOI = {10.1006/jmra.1993.1057},
  number = {3},
  journal = {Journal of Magnetic Resonance,  Series A},
  author = {Callaghan,  P.T. and Coy,  A. and Forde,  L.C. and Rofe,  C.J.},
  year = {1993},
  pages = {347–350}
}

@article{Stepisnik1999,
  title = {MRI Edge Enhancement as a Diffusive Discord of Spin Phase Structure},
  volume = {137},
  DOI = {10.1006/jmre.1998.1678},
  number = {1},
  journal = {Journal of Magnetic Resonance},
  author = {Stepišnik,  Janez and Duh,  Andrej and Mohorič,  Aleš and Serša,  Igor},
  year = {1999},
  pages = {154–160}
}

@article{Ozarslan2008,
  title = {Anisotropy Induced by Macroscopic Boundaries: Surface-Normal Mapping using Diffusion-Weighted Imaging},
  volume = {94},
  DOI = {10.1529/biophysj.107.124081},
  number = {7},
  journal = {Biophysical Journal},
  author = {\"{O}zarslan,  Evren and Nevo,  Uri and Basser,  Peter J.},
  year = {2008},
  pages = {2809–2818}
}

@article{Stoller1991,
  title = {Transverse spin relaxation in inhomogeneous magnetic fields},
  volume = {44},
  DOI = {10.1103/physreva.44.7459},
  number = {11},
  journal = {Physical Review A},
  author = {Stoller,  S. D. and Happer,  W. and Dyson,  F. J.},
  year = {1991},
  pages = {7459–7477}
}

@article{deSwiet1994,
  title = {Decay of nuclear magnetization by bounded diffusion in a constant field gradient},
  volume = {100},
  DOI = {10.1063/1.467127},
  number = {8},
  journal = {The Journal of Chemical Physics},
  author = {de Swiet,  Thomas M. and Sen,  Pabitra N.},
  year = {1994},
  pages = {5597–5604}
}

@article{deSwiet1995,
  title = {Diffusive Edge Enhancement in Imaging},
  volume = {109},
  DOI = {10.1006/jmrb.1995.1141},
  number = {1},
  journal = {Journal of Magnetic Resonance,  Series B},
  author = {Deswiet,  T.M.},
  year = {1995},
  pages = {12–18}
}

@article{Frohlich2006,
  title = {Effect of impermeable boundaries on diffusion-attenuated MR signal},
  volume = {179},
  DOI = {10.1016/j.jmr.2005.12.005},
  number = {2},
  journal = {Journal of Magnetic Resonance},
  author = {Frøhlich,  Astrid F. and Østergaard,  Leif and Kiselev,  Valerij G.},
  year = {2006},
  pages = {223–233}
}

@article{Hurlimann1995,
  title={Spin echoes in a constant gradient and in the presence of simple restriction},
  author={H{\"u}rlimann, M. D. and Helmer, K. G. and de Swiet, T. M. and Sen, P. N.},
  journal={Journal of Magnetic Resonance, Series A},
  volume={113},
  pages={260-264},
  year={1995},
  doi={10.1006/jmra.1995.1091}
}

@article{Grebenkov2018,
  title = {Diffusion MRI/NMR at high gradients: Challenges and perspectives},
  volume = {269},
  DOI = {10.1016/j.micromeso.2017.02.002},
  journal = {Microporous and Mesoporous Materials},
  author = {Grebenkov,  Denis S.},
  year = {2018},
  pages = {79–82}
}

@article{Moutal2019,
  title = {Localization regime in diffusion NMR: Theory and experiments},
  volume = {305},
  DOI = {10.1016/j.jmr.2019.06.016},
  journal = {Journal of Magnetic Resonance},
  author = {Moutal,  Nicolas and Demberg,  Kerstin and Grebenkov,  Denis S. and Kuder,  Tristan Anselm},
  year = {2019},
  pages = {162–174}
}

@article{Carr1954,
  title = {Effects of Diffusion on Free Precession in Nuclear Magnetic Resonance Experiments},
  volume = {94},
  DOI = {10.1103/physrev.94.630},
  number = {3},
  journal = {Physical Review},
  author = {Carr,  H. Y. and Purcell,  E. M.},
  year = {1954},
  pages = {630–638}
}

@article{Stejskal1965,
  title = {Spin Diffusion Measurements: Spin Echoes in the Presence of a Time-Dependent Field Gradient},
  volume = {42},
  DOI = {10.1063/1.1695690},
  number = {1},
  journal = {The Journal of Chemical Physics},
  author = {Stejskal,  E. O. and Tanner,  J. E.},
  year = {1965},
  pages = {288-292}
}

@article{Williamson2019,
  title = {Magnetic resonance measurements of cellular and sub-cellular membrane structures in live and fixed neural tissue},
  volume = {8},
  DOI = {10.7554/elife.51101},
  journal = {eLife},
  author = {Williamson,  Nathan H and Ravin,  Rea and Benjamini,  Dan and Merkle,  Hellmut and Falgairolle,  Melanie and O’Donovan,  Michael James and Blivis,  Dvir and Ide,  Dave and Cai,  Teddy X and Ghorashi,  Nima S and Bai,  Ruiliang and Basser,  Peter J},
  year = {2019}
}

@article{Williamson2025,
  title = {Passive water exchange between multiple sites can explain why apparent exchange rate constants depend on ionic and osmotic conditions in gray matter},
  DOI = {10.1016/j.mrl.2025.200225},
  journal = {Magnetic Resonance Letters},
  author = {Williamson,  Nathan H. and Ravin,  Rea and Cai,  Teddy X. and Rey,  Julian A. and Basser,  Peter J.},
  year = {2025},
  pages = {200225}
}

@article{Cheng2023,
  title = {Using deep learning to accelerate magnetic resonance measurements of molecular exchange},
  volume = {159},
  DOI = {10.1063/5.0159343},
  number = {5},
  journal = {The Journal of Chemical Physics},
  author = {Cheng,  Zhaowei and Hu,  Songtao and Han,  Guangxu and Fang,  Ke and Jin,  Xinyu and Ordinola,  Alfredo and \"{O}zarslan,  Evren and Bai,  Ruiliang},
  year = {2023}
}

@article{Herberthson2025,
  title = {Time-dependent diffusion in one-dimensional disordered media decorated by permeable membranes: Theoretical findings backed by simulations and a new disorder class},
  volume = {37},
  DOI = {10.1063/5.0272370},
  number = {6},
  journal = {Physics of Fluids},
  author = {Herberthson,  Magnus and Basser,  Peter J. and \"{O}zarslan,  Evren},
  year = {2025}
}

@article{Cai2025,
  title = {Measuring the velocity autocorrelation function using diffusion NMR},
  volume = {162},
  DOI = {10.1063/5.0258081},
  number = {17},
  journal = {The Journal of Chemical Physics},
  author = {Cai,  Teddy X. and Williamson,  Nathan H. and Ravin,  Rea and Herberthson,  Magnus and \"{O}zarslan,  Evren and Basser,  Peter J.},
  year = {2025}
}

@article{Callaghan1997,
  title = {A Simple Matrix Formalism for Spin Echo Analysis of Restricted Diffusion under Generalized Gradient Waveforms},
  volume = {129},
  DOI = {10.1006/jmre.1997.1233},
  number = {1},
  journal = {Journal of Magnetic Resonance},
  author = {Callaghan,  Paul T.},
  year = {1997},
  pages = {74–84}
}

@article{Caprihan1996,
  title = {A Multiple-Narrow-Pulse Approximation for Restricted Diffusion in a Time-Varying Field Gradient},
  volume = {118},
  DOI = {10.1006/jmra.1996.0013},
  number = {1},
  journal = {Journal of Magnetic Resonance,  Series A},
  author = {Caprihan,  A. and Wang,  L.Z. and Fukushima,  Eiichi},
  year = {1996},
  pages = {94–102}
}

@article{Barzykin1999,
  title = {Theory of Spin Echo in Restricted Geometries under a Step-wise Gradient Pulse Sequence},
  volume = {139},
  DOI = {10.1006/jmre.1999.1778},
  number = {2},
  journal = {Journal of Magnetic Resonance},
  author = {Barzykin,  A.V.},
  year = {1999},
  pages = {342–353}
}

@article{Grebenkov2007,
  title = {NMR survey of reflected Brownian motion},
  volume = {79},
  DOI = {10.1103/revmodphys.79.1077},
  number = {3},
  journal = {Reviews of Modern Physics},
  author = {Grebenkov,  Denis S.},
  year = {2007},
  pages = {1077–1137}
}

@article{Grebenkov2008,
  title = {Laplacian eigenfunctions in NMR. I. A numerical tool},
  volume = {32A},
  DOI = {10.1002/cmr.a.20117},
  number = {4},
  journal = {Concepts in Magnetic Resonance Part A},
  author = {Grebenkov,  Denis S.},
  year = {2008},
  pages = {277–301}
}

@article{Herberthson2017,
  title = {Dynamics of local magnetization in the eigenbasis of the Bloch-Torrey operator},
  volume = {146},
  DOI = {10.1063/1.4978621},
  number = {12},
  journal = {The Journal of Chemical Physics},
  author = {Herberthson,  Magnus and \"{O}zarslan,  Evren and Knutsson,  Hans and Westin,  Carl-Fredrik},
  year = {2017}
}

@article{Zientara1980,
  title = {Spin-echoes for diffusion in bounded, heterogeneous media: A numerical study},
  volume = {72},
  DOI = {10.1063/1.439190},
  number = {2},
  journal = {The Journal of Chemical Physics},
  author = {Zientara,  Gary P. and Freed,  Jack H.},
  year = {1980},
  pages = {1285–1292}
}

@article{Blees1994,
  title = {The Effect of Finite Duration of Gradient Pulses on the Pulsed-Field-Gradient NMR Method for Studying Restricted Diffusion},
  volume = {109},
  DOI = {10.1006/jmra.1994.1156},
  number = {2},
  journal = {Journal of Magnetic Resonance,  Series A},
  publisher = {Elsevier BV},
  author = {Blees,  M.H.},
  year = {1994},
  pages = {203–209}
}

@article{Sen1999,
  title = {Spin echoes of nuclear magnetization diffusing in a constant magnetic field gradient and in a restricted geometry},
  volume = {111},
  DOI = {10.1063/1.480009},
  number = {14},
  journal = {The Journal of Chemical Physics},
  author = {Sen,  Pabitra N. and André,  Axel and Axelrod,  Scott},
  year = {1999},
  pages = {6548–6555}
}

@article{Salikhov1996,
  title = {A theoretical approach to the analysis of arbitrary pulses in magnetic resonance},
  volume = {262},
  DOI = {10.1016/0009-2614(96)01044-5},
  number = {1–2},
  journal = {Chemical Physics Letters},
  author = {Salikhov,  Kev M. and Schneider,  David J. and Saxena,  Sunil and Freed,  Jack H.},
  year = {1996},
  pages = {17–26}
}

@article{Crank1947,
  title = {A practical method for numerical evaluation of solutions of partial differential equations of the heat-conduction type},
  volume = {43},
  DOI = {10.1017/s0305004100023197},
  number = {1},
  journal = {Mathematical Proceedings of the Cambridge Philosophical Society},
  author = {Crank,  J. and Nicolson,  P.},
  year = {1947},
  pages = {50–67}
}

@article{Trotter1959,
  title = {On the product of semi-groups of operators},
  volume = {10},
  DOI = {10.1090/s0002-9939-1959-0108732-6},
  number = {4},
  journal = {Proceedings of the American Mathematical Society},
  author = {Trotter,  H. F.},
  year = {1959},
  pages = {545–551}
}

@article{Neuman1974,
  title = {Spin echo of spins diffusing in a bounded medium},
  volume = {60},
  DOI = {10.1063/1.1680931},
  number = {11},
  journal = {The Journal of Chemical Physics},
  author = {Neuman,  C. H.},
  year = {1974},
  pages = {4508–4511}
}

@article{LeBihan1986,
  title = {MR imaging of intravoxel incoherent motions: application to diffusion and perfusion in neurologic disorders.},
  volume = {161},
  DOI = {10.1148/radiology.161.2.3763909},
  number = {2},
  journal = {Radiology},
  author = {Le Bihan,  D and Breton,  E and Lallemand,  D and Grenier,  P and Cabanis,  E and Laval-Jeantet,  M},
  year = {1986},
  pages = {401–407}
}

@BOOK{Casanova2011,
  title     = "{Single-Sided} {NMR}",
  editor    = "Casanova, Federico and Perlo, Juan and Blumich, Bernhard",
  publisher = "Springer",
  edition   =  2011,
  year      =  2011,
  address   = "Berlin, Germany"
}

@article{Brownstein1979,
  title = {Importance of classical diffusion in NMR studies of water in biological cells},
  volume = {19},
  DOI = {10.1103/physreva.19.2446},
  number = {6},
  journal = {Physical Review A},
  author = {Brownstein,  K. R. and Tarr,  C. E.},
  year = {1979},
  pages = {2446–2453}
}

@article{Afrough2024,
  title = {Magnetic Resonance Relaxation in Heterogeneous Materials is Analogous to First-Order Chemical Reaction},
  volume = {151},
  DOI = {10.1007/s11242-024-02075-y},
  number = {7},
  journal = {Transport in Porous Media},
  author = {Afrough,  Armin},
  year = {2024},
  pages = {1493–1509}
}

@article{Grebenkov2014,
  title = {Exploring diffusion across permeable barriers at high gradients. II. Localization regime},
  volume = {248},
  DOI = {10.1016/j.jmr.2014.08.016},
  journal = {Journal of Magnetic Resonance},
  author = {Grebenkov,  Denis S.},
  year = {2014},
  pages = {164–176}
}

@article{Lee1993,
  title = {Two-dimensional inverse Laplace transform NMR: altered relaxation times allow detection of exchange correlation},
  volume = {115},
  DOI = {10.1021/ja00070a022},
  number = {17},
  journal = {Journal of the American Chemical Society},
  author = {Lee,  Jing Huei and Labadie,  Christian and Springer,  Charles S. and Harbison,  Gerard S.},
  year = {1993},
  pages = {7761–7764}
}

@article{Washburn2006,
  title = {Tracking Pore to Pore Exchange Using Relaxation Exchange Spectroscopy},
  volume = {97},
  DOI = {10.1103/physrevlett.97.175502},
  number = {17},
  journal = {Physical Review Letters},
  author = {Washburn,  K. E. and Callaghan,  P. T.},
  year = {2006}
}

@article{Ordinola2024,
  title = {Limitations and generalizations of the first order kinetics reaction expression for modeling diffusion-driven exchange: Implications on NMR exchange measurements},
  volume = {160},
  DOI = {10.1063/5.0188865},
  number = {8},
  journal = {The Journal of Chemical Physics},
  author = {Ordinola,  Alfredo and \"{O}zarslan,  Evren and Bai,  Ruiliang and Herberthson,  Magnus},
  year = {2024}
}

@inproceedings{Lee2023,
title = {Localization Regime of Diffusion (LoRD) in mouse and pig cortical gray matter and its sensitivity to soma size},
author = {Lee, Hong-Hsi and Wang, Nian and Fieremans, Els and Huang, Susie Y and Novikov, Dmitry S},
booktitle = {Proceedings of the 31st Annual Meeting of the International Society of Magnetic Resonance in Medicine (ISMRM)},
year = {2023},
address = {Toronto, Canada},
pages = {0681}
}

@article{Laun2011,
  title = {Determination of the Defining Boundary in Nuclear Magnetic Resonance Diffusion Experiments},
  volume = {107},
  DOI = {10.1103/physrevlett.107.048102},
  number = {4},
  journal = {Physical Review Letters},
  author = {Laun,  Frederik Bernd and Kuder,  Tristan Anselm and Semmler,  Wolfhard and Stieltjes,  Bram},
  year = {2011}
}

@article{Kac1966,
  title = {Can One Hear the Shape of a Drum?},
  volume = {73},
  DOI = {10.2307/2313748},
  number = {4},
  journal = {The American Mathematical Monthly},
  author = {Kac,  Mark},
  year = {1966},
  pages = {1}
}

@article{Song2000,
  title = {Detection of the High Eigenmodes of Spin Diffusion in Porous Media},
  volume = {85},
  DOI = {10.1103/physrevlett.85.3878},
  number = {18},
  journal = {Physical Review Letters},
  author = {Song,  Yi-Qiao},
  year = {2000},
  pages = {3878–3881}
}

@article{Callaghan1995,
  title = {Frequency-Domain Analysis of Spin Motion Using Modulated-Gradient NMR},
  volume = {117},
  DOI = {10.1006/jmra.1995.9959},
  number = {1},
  journal = {Journal of Magnetic Resonance,  Series A},
  author = {Callaghan,  Paul T. and Stepišnik,  Janes},
  year = {1995},
  pages = {118–122}
}

@article{Parsons2005,
  title = {Temporal diffusion spectroscopy: Theory and implementation in restricted systems using oscillating gradients},
  volume = {55},
  DOI = {10.1002/mrm.20732},
  number = {1},
  journal = {Magnetic Resonance in Medicine},
  author = {Parsons,  Edward C. and Does,  Mark D. and Gore,  John C.},
  year = {2005},
  pages = {75–84}
}

@article{Ordinola2023,
  title = {On the sampling strategies and models for measuring diffusion exchange with a double diffusion encoding sequence},
  volume = {3},
  DOI = {10.1016/j.mrl.2023.05.003},
  number = {3},
  journal = {Magnetic Resonance Letters},
  author = {Ordinola,  Alfredo and Cai,  Shan and Lundberg,  Peter and Bai,  Ruiliang and \"{O}zarslan,  Evren},
  year = {2023},
  pages = {232–247}
}

@inproceedings{Hansen2000,
title = "The L-curve and its use in the numerical treatment of inverse problems",
author = "Hansen, \{Per Christian\}",
year = "2000",
language = "English",
booktitle = "InviteComputational Inverse Problems in Electrocardiology",
publisher = "WIT Press",
pages = {119-142}
}

@article{Song2016,
  title = {The robust identification of exchange from T2–T2 time-domain features},
  volume = {265},
  DOI = {10.1016/j.jmr.2016.02.001},
  journal = {Journal of Magnetic Resonance},
  author = {Song,  Ruobing and Song,  Yi-Qiao and Vembusubramanian,  Muthusamy and Paulsen,  Jeffrey L.},
  year = {2016},
  pages = {164–171}
}

\end{document}